\documentclass[10pt,aps,prb,twocolumn,amsmath,amssymb,floats,showpacs,floatfix,superscriptaddress]{revtex4-2}
\usepackage{graphicx}
\usepackage{bm}
\usepackage{amsmath}
\usepackage{amsfonts}
\usepackage{amsbsy}
\usepackage{physics}
\usepackage{verbatim}
\usepackage{color}
\usepackage{soul}
\usepackage{indentfirst}
\usepackage{upgreek}
\usepackage{placeins}
\usepackage{ragged2e}
\usepackage[
    urlcolor=blue,
    colorlinks=true,
    citecolor=blue,
    linkcolor=blue
]{hyperref}

\usepackage[nameinlink]{cleveref}

\crefmultiformat{figure}
    {~#2#1#3}
    {,\!~#2#1#3}
    {, #2#1#3}
    { and~#2#1#3}

\crefname{equation}{Eq.}{Eqs.}
\crefname{figure}{Fig.}{Figs.}
\crefname{section}{Sec.}{Secs.}
\crefname{appendix}{Appendix}{Appendices}

\usepackage{orcidlink}
\newcommand{\orcidauthorA}{%
  \protect\orcidlink{0000-0002-1207-0407}%
}
\newcommand{\orcidauthorB}{%
  \protect\orcidlink{0009-0007-6239-9407}%
}
\newcommand{\orcidauthorC}{%
  \protect\orcidlink{0000-0002-1105-3277}%
}
\newcommand{\orcidauthorD}{%
  \protect\orcidlink{0000-0003-2590-6231}%
}
\newcommand{\orcidauthorE}{%
  \protect\orcidlink{0000-0003-0839-6268}%
}

\begin{document}
 \newcommand{\WARSAW}{\affiliation{International Research Centre MagTop, Institute of Physics, Polish Academy of Sciences, Aleja Lotnikow 32/46, PL-02668 Warsaw, Poland}}
 \newcommand{\HAMBURG}{\affiliation{The Hamburg Centre for Ultrafast Imaging, Luruper Chaussee 149, 22761 Hamburg, Germany}}
\newcommand{\UHH}{\affiliation{I. Institut für Theoretische Physik, Universität Hamburg, Notkestra{\ss}e 9, 22607 Hamburg, Germany}}
\newcommand{\INDIA}{\affiliation{Theoretical Physics Division, Physical Research Laboratory, Ahmedabad, 380009, India}}
\newcommand{\Basel}{\affiliation{Department of Physics, University of Basel, Klingelbergstrasse 82, 4056 Basel, Switzerland}}
\newcommand{\BRAUNSCHWEIG}{\affiliation{Institut für Mathematische Physik, Technische Universität Braunschweig, D-38106 Braunschweig, Germany}}

\newcommand{\COFIRST}{\thanks{DD and II contributed equally to this work.}}
\newcommand{\COFIRSTTEXT}{DD and II contributed equally to this work.}

\title{Supercurrent detection and manipulation of topological phase transitions in Shiba–Majorana hybrid systems}

\author{Debika Debnath\orcidauthorA}
\email[e-mail:]{debika.uoh@gmail.com}
\thanks{\\A major part of this work was carried out while D.D. was affiliated with Physical Research Laboratory, India.}
\INDIA
\BRAUNSCHWEIG
\altaffiliation{DD and II contributed equally to this work}

\author{Ioannis Ioannidis\orcidauthorB}
\email[e-mail:]{ioannis.ioannidis@uni-hamburg.de}
\HAMBURG
\UHH
\altaffiliation{DD and II contributed equally to this work}

\author{Paramita Dutta\orcidauthorD}
\INDIA

\author{Mircea Trif\orcidauthorC}
\WARSAW
\Basel

\author{Thore Posske\orcidauthorE}
\HAMBURG
\UHH

\begin{abstract}

The non-Abelian statistics of Majorana zero modes has inspired numerous proposals for their detection and manipulation in topological superconductors. Implementations based on magnetic adatoms deposited on superconductors draw particular attention due to their capabilities for precise atomic manipulation and the control over disorder. Here, we propose a scheme for detecting changes in the ground state parity of a topologically non-trivial adatom system by passing supercurrent through their low-energy modes. We unravel characteristic discontinuities in the critical current driven by zero-energy level crossings. We apply these findings to a setup where the Majorana coupling is mediated by a single control magnetic adatom hosting a Yu-Shiba-Rusinov state, and test the robustness of our results against finite temperatures and different tunneling regimes. Our findings introduce a non-invasive approach for reading out and controlling the ground state parity of Majorana states in Shiba-Majorana hybrid systems.                                                                                 
\end{abstract}
\maketitle
\section{INTRODUCTION}\label{Introduction}

Majorana zero modes and their non-Abelian statistics potentially offer the possibility for building qubits with topological protection ~\cite{Alicea2011, Sau2011, Sarma2015, Li2016, Orien2018, Mascot2023, Liu2023b, Read2000, Ivanov2001, Fu2008} and thereby serve as a starting point for topological quantum computing~\cite{Nayak2008, Aasen2016}. Although substantial experimental effort has reported signatures consistent with Majorana zero modes in various hybrid magnet–superconductor systems, their unambiguous identification and controlled manipulation remain ongoing challenges~\cite{Leijnse2012, Mourik2012, Perge2014, Kim2018, Morales2019, Mishra2021, Zhang2021, Liebhaber2022, Schneider2022, Aghaee2023, Aghaee2025, Frolov2026}. In general, performing fusion or braiding protocols requires control over the coupling between Majorana modes~\cite{vanHeck2012, Liu2023, Yu2025}. In the semiconductor/superconductor nanowire platform, this control can be achieved by applying electrostatic potentials~\cite{Luna2024}, while for magnetic adatom chains that induce hybridizing Yu-Shiba-Rusinov~\cite{Yu1965Bound, Shiba1968classical, Rusinov1969theory} states on superconducting surfaces --- referred to as Shiba states here --- this control may be achieved by altering the local magnetic field of a single adatom~\cite{Chakraborty2023, Ohnmacht2023, Awoga2024} using electron-spin-resonance scanning tunneling microscopy (ESR-STM) techniques~\cite{Baumann2015, Yang2019}. Non-invasive access to the ground-state parity of a Majorana qubit, without inducing quasiparticle interference or parity flips~\cite{Diego2012}, can be facilitated by reading out the ground state energy, which can be achieved by transport measurements in superconducting systems in the context of Josephson tunneling spectroscopy~\cite{Jeon2017}. In such systems, a finite phase difference between superconductors induces tunneling processes of Cooper pairs and generates a dissipationless Josephson supercurrent~\cite{Josephson1962}. When the spin-flipping dominates over the spin-conserving part of the tunneling processes, a $\pi$ phase shift is introduced in the current-phase relation~\cite{Kulik1965}. This effect has also been established in correlated and multi-orbital quantum dot Josephson junctions, where the supercurrent serves as a detection mechanism for the ground state occupancy of the quantum dot~\cite{Vecino2003, vanDam2006, Droste2012, Delagrange2016, Hsu2020, Debbarma2022}. Moreover, the characteristic $\pi$ phase shift has inspired proposals for realizing qubits with large decoherence times and consequently for building complex quantum circuits~\cite{Ioffe1999, Yamashita2005, Feofanov2010, Gingrich2016, Kim2024}, distinguishing non-trivial topological excitations in nanowire setups~\cite{Awoga2019} and measuring fermion parity in the quantum spin Hall effect~\cite{Beenakker2013}. 

In superconducting junctions that include Shiba states, $0-\pi$ phase transitions have been theoretically predicted to accompany changes in magnetic coupling and orientation~\cite{Karan2022, Chakraborty2023}. Motivated by these considerations, we explore adatom-based architectures and propose a non-invasive, current-driven detection protocol to identify topological quantum phase transitions in Shiba-Majorana junction. 
Let us review the general mechanism. At zero temperature, the supercurrent in a Josephson junction is determined by~\cite{Beenakker2013, Probst2016}
\begin{equation}\label{eq:SupercurrentExpressionZeroT}
  \mathcal{I}(\phi)=2\frac{\partial E_{\mathrm{GS}}(\phi)}{\partial \phi},
\end{equation}
where the ground state energy $E_\mathrm{GS}(\phi)$ depends on the superconducting phase $\phi$ and is constructed within the single-particle picture as $E_\mathrm{GS}(\phi)=-\sum_{n}|\epsilon_n(\phi)|/2$, where $n$ labels the phase dependent single-particle eigenvalues $\epsilon_n(\phi)$. Here, we set $e = \hbar = 1$. 
At a quantum phase transition, a subgap state crosses zero energy and the derivative of ground state energy exhibits jumps when changing $\phi$, assuming that relaxation processes enable transitions between fermion parity sectors such that the system follows the lowest-energy branch~\cite{Beenakker2013, Probst2016}. This generic feature of quantum phase transitions does not depend on the microscopic details of the system~\cite{Vecino2003, Huang2020, Karan2022}. 
From \cref{eq:SupercurrentExpressionZeroT}, it is evident that the resulting supercurrent discontinuity stems from $\epsilon_n(\phi)=0$ for some eigenstate labeled by $n$  and contributions from higher-energy states can be neglected while identifying ground-state parity changes. As a central observable for the supercurrent jumps, we calculate the critical current~\cite{Beenakker1992}
\begin{align}
\mathcal{I}_\mathrm{c}=\max_{\phi} [\mathcal{I}(\phi)]\,.
\end{align}
When the tip–system coupling remains weak compared with the s-wave superconducting gap, the resulting hybridization can substantially modify the low-energy subgap spectrum. In particular, it can induce zero-energy crossings and thereby change the ground-state parity~\cite{Beenakker2013, Rodero2011,Vecino2003, Zitko2015, Karan2022}, without relying on being close to a quantum phase transition in parameter space. 

Here, we study a Josephson junction involving hybridized Majorana and Shiba states, focusing on how the superconducting phase bias modifies the low-energy spectrum and the resulting supercurrent.
The spatial features of the wavefunctions of the Shiba and Majorana states physically determine the tunneling energy scales and ultimately the supercurrent characteristics. We present the supercurrent jump as a controlled readout mechanism for detecting topological quantum phase transitions in our setup. We also assess the impact of experimental constraints, for example, finite temperature, the broadening of the superconducting leads, and particle-hole asymmetry on our proposed detection scheme and describe how the setup can be implemented in a current-biased STM setup~\cite{ScienceMartinis88}. 

The remainder of this manuscript is organized as follows: In \cref{Model}, we introduce the model system and the formalism for calculating the supercurrent in a Josephson junction. \cref{Results} presents our main results, where we analyze the supercurrent signatures of the Majorana quantum phase transition at different tunneling rate regimes and the effect of finite temperature on the phase transition. We also present the effects of experimental non-idealities, including the tip-Majorana coupling and dissipation in \cref{TipMajoranaCoupling}. In \cref{Implementation}, we discuss the physical implementation analog of supercurrent detection to a quantum state. Finally, we summarize our results and provide an outlook in \cref{Conclusion}. As a foundation of our setup, we provide an analytical description of the parity dependent Josephson current for a single Shiba state in Appendix~\ref{Appendix:CurrentforShiba}.
\section{Model and Formalism} \label{Model}
\begin{figure}[t]
    \centering
    \includegraphics[width=0.98\linewidth]{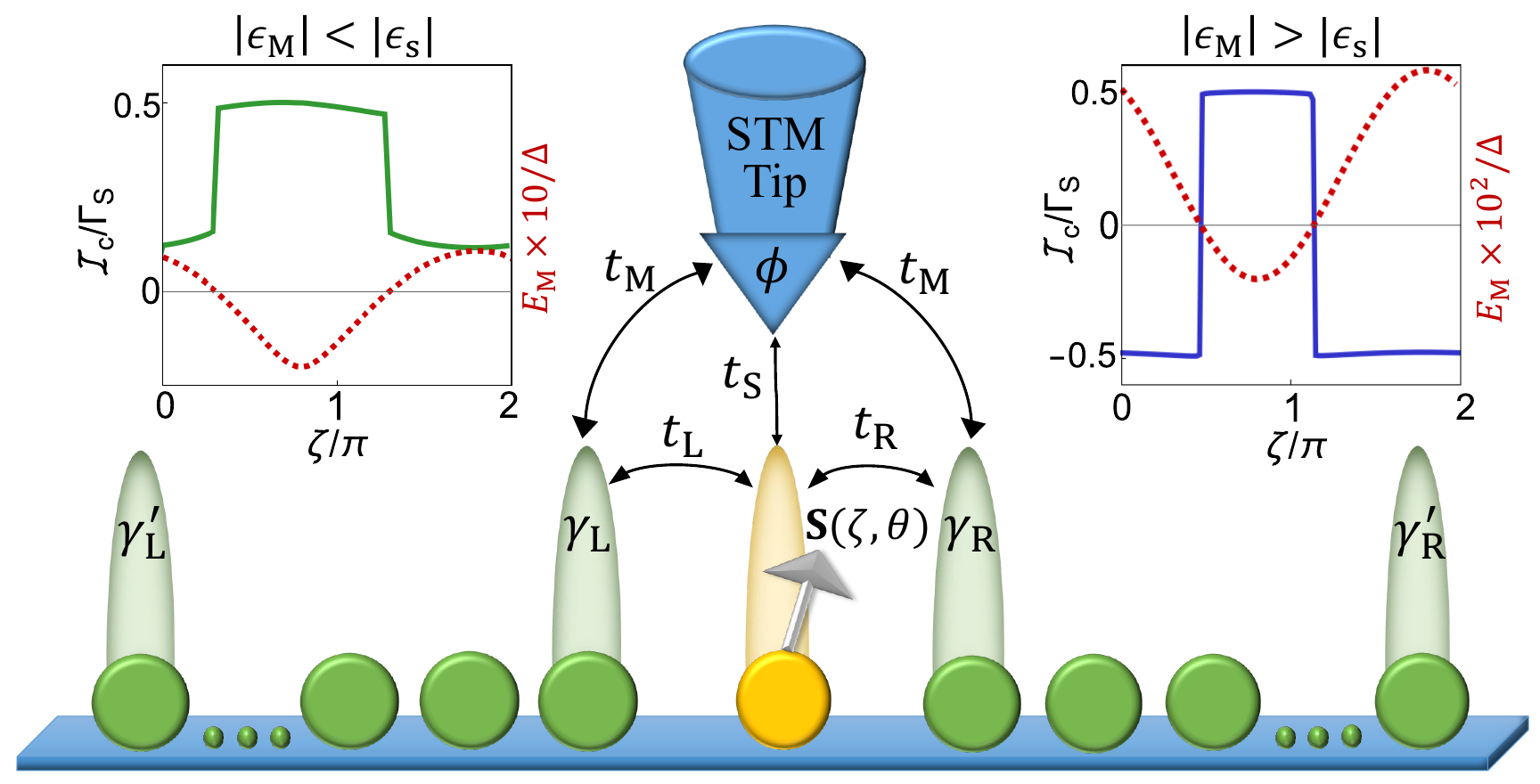}
    \caption{Schematic setup for supercurrent-based detection and manipulation of the Majorana hybridization mediated by a control Shiba state. Two chains of magnetic adatoms (green) are deposited on a superconducting substrate (blue). In the nontrivial topological phase, each chain supports a pair of Majoranas, labeled $\gamma_{\rm L(R)}$ and $\gamma'_{\rm L(R)}$. The bare hybridization energy associated with the inner Majoranas, $\gamma_{\rm L}$ and $\gamma_{\rm R}$, is denoted by $\epsilon_{\mathrm M}$. A control magnetic adatom (yellow), with bare Shiba energy $\epsilon_{\mathrm S}$ and magnetic moment $\mathbf S=S(\sin\theta\cos\upzeta,\sin\theta\sin\upzeta,\cos\theta)$, parametrized by the azimuthal and polar angles $(\upzeta,\theta)$, is placed between the chains. The hybridization of the inner Majoranas is mediated by the control Shiba state through the couplings $t_{\rm  L(R)}$, which renormalize the Majorana energy to $E_{\mathrm M}$, see \cref{eq:AppendixPoles}. A superconducting STM tip with a phase bias $\phi$ is tunnel-coupled to the control Shiba state and to the Majorana sector with the (complex) tunneling amplitudes $t_{\rm S}$ and $t_{\rm M}$, respectively. The insets show the supercurrent through the control Shiba state for the two regimes $|\epsilon_{\mathrm M}|\gtrless|\epsilon_{\mathrm S}|$. The critical angles at which the supercurrent exhibits a jump coincide with the sign change of the inner Majoranas hybridization energy $E_{\mathrm M}$, indicated on the right axis and marked by the red dashed lines.}
    \label{fig:Schematics}
\end{figure}
We consider a concrete realization in which Majorana edge modes emerge at the ends of two magnetic-adatom chains. The magnetic impurities induce spin-polarized Shiba bands inside the gap of the superconducting substrate, which can enter a nontrivial topological phase. An additional control adatom modifies the coupling between the two inner Majorana modes through indirect tunneling processes mediated by the common superconducting substrate ~\cite{Fan2021,Awoga2024}, see \cref{fig:Schematics}. By changing the magnetic orientation of the control adatom, one can tune its Shiba-state hybridization with the Majorana modes~\cite{Awoga2024}. Our aim is to probe this tunable hybridization through superconducting STM measurements of the supercurrent. To connect the microscopic STM setup to the low-energy description used in the following, we first decompose the full model Hamiltonian as 
\begin{equation} 
H = H_{\rm tip} + H_{\rm sub} + H_{\rm tip-sub}\,, 
\label{eq:Hmicro_decomposition} 
\end{equation} 
where \(H_{\rm tip}\) describes the superconducting STM tip, \(H_{\rm sub}\) contains the superconducting substrate, the magnetic chains, and the control adatom, and \(H_{\rm tip-sub}\) accounts for local electron tunneling between the tip and the substrate. The microscopic tunneling Hamiltonian is~\cite{Pientka2013, inbook}
\begin{equation} 
H_{\rm tip-sub} = t_{\rm tip} \sum_{\mathbf{k},\sigma} \left[ e^{-i\phi/2} a_{\mathbf{k}\sigma}^{\dagger} \psi_{\sigma}(\mathbf{r}_{\rm tip}) + {\rm H.c.} \right], 
\label{eq:Htun_micro} 
\end{equation} 
where \(a_{\mathbf{k}\sigma}\) annihilates an electron in the superconducting tip, \(\psi_{\sigma}(\mathbf{r}_{\rm tip})\) is the substrate electron field evaluated at the tip position, \(t_{\rm tip}\) is the microscopic tip--substrate tunneling amplitude, and \(\phi\) is the superconducting phase difference between the tip and the substrate. Since we focus on the contribution of the discrete in-gap states to the supercurrent, we project $H$ onto the low-energy subspace spanned by the Shiba state localized on the control adatom and the two inner Majorana modes. Defining the corresponding projector by \(P_{\rm low}\), the substrate electron field at the tip position becomes \cite{RubyPRL15}
\begin{equation}  
P_{\rm low} \psi_{\sigma}(\mathbf{r}_{\rm tip}) P_{\rm low} = u_{{\rm S}\sigma}d + v_{{\rm S}\sigma}^{*}d^{\dagger} + u_{{\rm M}\sigma}c + v_{{\rm M}\sigma}^{*}c^{\dagger}, 
\label{eq:projected_field} 
\end{equation} 
where \(d\) annihilates the control-Shiba state, while $c=(\gamma_{\rm L}+i\gamma_{\rm R})/2$ is the fermionic mode formed from the two inner Majorana operators. The coefficients \(u_{{\rm S}\sigma}\) and \(v_{{\rm S}\sigma}\) are the local electron and hole components of the control-Shiba wave function, while \(u_{{\rm M}\sigma}\) and \(v_{{\rm M}\sigma}\) are the corresponding, generally complex, components of the Majorana sector. Substituting \cref{eq:projected_field} into \cref{eq:Htun_micro}, and projecting \( H_{\rm low} = P_{\rm low} H_{\rm sub} P_{\rm low}\) onto the same low-energy subspace, which describes the retained Shiba-Majorana subspace, 
yields  $H_{\rm eff} = H_{\rm tip} + H_{\rm low} + H_{\rm tip-low}$ with 
\begin{align} 
H_{\rm tip} &= \sum_{\mathbf{k},\sigma} \epsilon_{\mathbf{k}\sigma} a_{\mathbf{k}\sigma}^{\dagger} a_{\mathbf{k}\sigma} + \Delta \sum_{\mathbf{k}} \left( a_{\mathbf{k}\uparrow}^{\dagger} a_{-\mathbf{k}\downarrow}^{\dagger} + a_{-\mathbf{k}\downarrow} a_{\mathbf{k}\uparrow} \right), \nonumber\\ 
H_{\rm low} &= \epsilon_{\rm M}c^{\dagger}c + \epsilon_{\rm S}d^{\dagger}d + \left( t_{1}d^{\dagger}c + t_{2}dc + {\rm H.c.} \right), \nonumber\\ 
H_{\rm tip-low} &= t_{\rm S}e^{i\phi_{1}/2} \sum_{\mathbf{k}} \left( u\,d^{\dagger}a_{\mathbf{k}\uparrow} + h\,d\,a_{\mathbf{k}\downarrow} \right) \nonumber\\ &\quad + t_{\rm M}e^{i\phi_{2}/2} \sum_{\mathbf{k},\sigma}a_{\mathbf{k}\sigma}\left(f_{\sigma}c^{\dagger} + g_{\sigma}c\right) + {\rm H.c.}\,. 
\label{eq:HamiltonianLowEenergy} 
\end{align} 
Here, \(H_{\rm tip}\) describes an \(s\)-wave superconducting STM tip with order parameter \(\Delta\) and normal-state dispersion \(\epsilon_{\mathbf{k}\sigma}\). The parameters \(\epsilon_{\rm S}\) and \(\epsilon_{\rm M}\) denote the bare energies of the control-Shiba and Majorana fermionic modes, respectively, while \(t_{1}\) and \(t_{2}\) describe their normal and anomalous hybridization. Because these couplings depend on the magnetic orientation of the control adatom, they provide the control mechanism of the device~\cite{Awoga2024}. The effective amplitudes \(t_{\rm S}\) and \(t_{\rm M}\) originate from the same microscopic tip--substrate tunneling matrix element \(t_{\rm tip}\), but are weighted by the local Shiba and Majorana wave functions at the tip position, respectively. Defining the corresponding local Nambu weights as 
\begin{equation} 
W_{\lambda} = \sum_{\sigma} \left( |u_{\lambda\sigma}|^{2} + |v_{\lambda\sigma}|^{2} \right), \qquad \lambda={\rm S,M}\,, \label{eq:local_weights} 
\end{equation} 
the projected tunneling amplitudes are $t_{\rm S, M} = t_{\rm tip}\sqrt{W_{\rm S, M}}$ (see Appendix \ref{Appendix:CurrentforShiba}). The normalized particle and hole components of the control-Shiba state that enter \(H_{\rm tip-low}\) are given by 
\(u = u_{{\rm S}\uparrow}/\sqrt{W_{\rm S}}\) and \(h = v^*_{{\rm S}\downarrow}/\sqrt{W_{\rm S}}\), which generally satisfy \(|u|^{2} + |h|^{2} = 1\) for the spin convention used above, where
the retained Shiba components are $u_{{\rm S}\uparrow}$ and
$v_{{\rm S}\downarrow}$, while
$u_{{\rm S}\downarrow}=v_{{\rm S}\uparrow}=0$.
We restrict our analysis to a Shiba state with equal particle and hole weight $(u=h)$ and use a classical spin configuration. Similarly, the normalized particle and hole components of the fermionic Majorana-sector mode \(H_{\rm tip-low}\) are \(f_\sigma = u_{{\rm M}\sigma}/\sqrt{W_{\rm M}}\) and \(g_\sigma = v^*_{{\rm M}\sigma}/\sqrt{W_{\rm M}}\), obeying \(\sum_\sigma (|f_\sigma|^{2} + |g_\sigma|^{2}) = 1\). Because the STM tip is positioned above the control adatom, \(W_{\rm S}\) is set by the local weight of the control-Shiba state. By contrast, \(W_{\rm M}\) probes the tail of the Majorana wave function at a distance \(R\) from the end of the topological chain, where the test adatom is positioned. Asymptotically, \(W_{\rm M}\propto e^{-2R/\xi_{\rm top}}\), up to oscillatory, algebraic, and spin-overlap prefactors \cite{Awoga2024}, where \(\xi_{\rm top}\simeq v_{\rm top}/\Delta_{\rm top}\) is the Majorana localization length. Here, \(v_{\rm top}\) denotes the effective Fermi velocity of the topological Shiba band, while \(\Delta_{\rm top}\) is the topological minigap opened at the Fermi points of this band. Consequently,
\begin{equation}
\frac{t_{\rm M}}{t_{\rm S}}
\propto
e^{-R/\xi_{\rm top}}\,.
\label{eq:tM_tS_scaling}
\end{equation}
This minigap is determined by the hybridization of the single-impurity Shiba states, the magnetic texture of the chain, spin-orbit coupling in the substrate, and the position of the Shiba band relative to the Fermi level \cite{Pientka2013}. In atomic Shiba-chain experiments, inferred topological minigaps in the tens of \(\mu{\rm eV}\) range have been reported. For example, fits to atomically constructed Mn chains on Nb(110) estimated a gap of order
\(50~\mu{\rm eV}\) \cite{Schneider2021, Schneider2022}, while stronger substrate spin-orbit coupling can enhance the minigap relative to the parent gap.

This separation of energy scales naturally realizes the regime $t_{\rm M}\ll t_{\rm S}$, in which the STM tip predominantly probes the Majorana modes indirectly through the controllable Shiba–Majorana hybridization rather than through direct tunneling. We take \(t_{\rm S}\) and \(t_{\rm M}\) to be real and absorb their intrinsic complex phases into \(\phi_{1}\) and \(\phi_{2}\). Only one combination of these phases corresponds to the externally imposed superconducting phase bias. Therefore, it is convenient to write \(\phi_{1}=\phi\) and \(\phi_{2}=\phi+\Delta\phi\), where \(\Delta\phi\) is a phase-independent relative phase between the Shiba and Majorana tunneling paths. This phase originates from the local complex structure of the corresponding in-gap wave functions at the position of the STM tip.

\begin{figure*}[t]
    \centering
    \includegraphics[width=1\linewidth,height=0.25\textwidth]{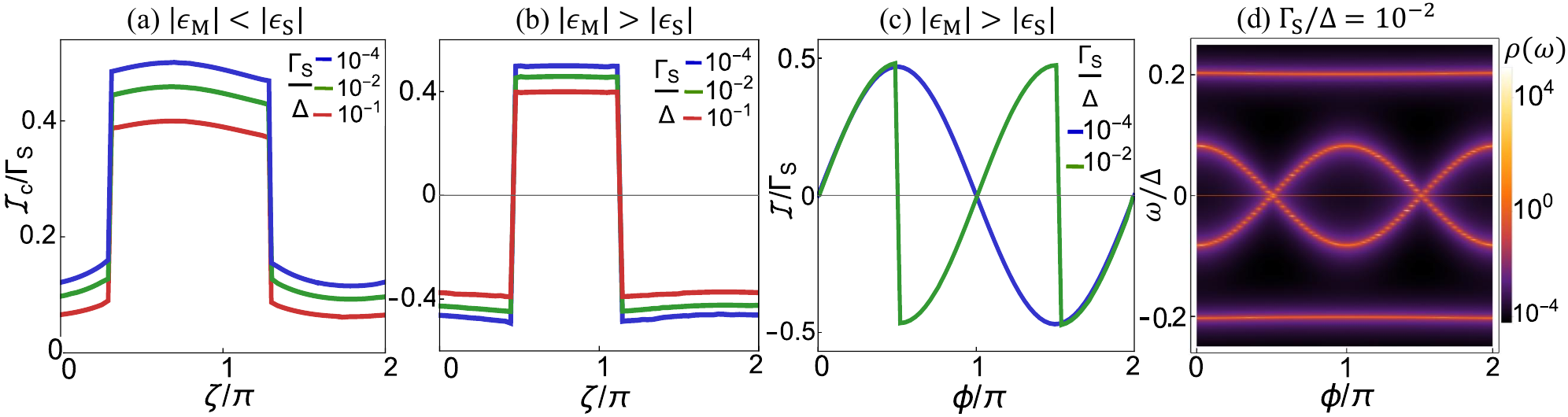}
    \caption{Supercurrent from the superconducting tip to the control magnetic adatom as a function of the adatom azimuthal angle $\upzeta$ and the superconducting phase bias $\phi$. 
    (a) Critical supercurrent as a function of $\upzeta$ for the regime $|\epsilon_{\mathrm M}|<|\epsilon_{\mathrm S}|$, with $\epsilon_{\mathrm M}=-0.02\Delta$ and $\epsilon_{\mathrm S}=0.04\Delta$, for several tip--Shiba tunneling rates $\Gamma_{\mathrm S}/\Delta$. 
    (b) Same as in (a), but for $|\epsilon_{\mathrm M}|>|\epsilon_{\mathrm S}|$, with $\epsilon_{\mathrm M}=-0.2\Delta$ and $\epsilon_{\mathrm S}=0.002\Delta$. 
    In both regimes, the discontinuities in the critical current occur at the critical magnetic orientations where the effective Majorana hybridization energy $E_{\mathrm M}$ changes sign, corresponding to the transition between the two parity/topological sectors shown in the inset panels of \cref{fig:Schematics}. We use $\theta=\pi/2$ in panels (a) and (b). 
    (c) Supercurrent as a function of the superconducting phase bias $\phi$ for different values of $\Gamma_{\mathrm S}/\Delta$, using $\epsilon_{\mathrm M}=-0.2\Delta$, $\epsilon_{\mathrm S}=0.002\Delta$, and $(\upzeta, \theta)=(\pi/2,\pi/2)$. At stronger tip--Shiba coupling, $\Gamma_{\mathrm S}/\Delta= 10^{-2}$ the current-phase relation develops a discontinuity associated with a zero-energy crossing. 
    (d) Corresponding density of states $\rho(\omega)$ for $\Gamma_{\mathrm S}/\Delta=10^{-2}$, showing the zero-energy crossing responsible for the current jump in panel (c). Here, we consider $k_B T/\Delta=10^{-5}$. Unless stated otherwise, the remaining parameters are the same as in panel (b). }
    \label{fig:2}
\end{figure*}
The normal and anomalous Shiba--Majorana hybridizations appearing in \(H_{\rm low}\) are related to the couplings of the control-Shiba state to the two inner Majorana modes according to $t_{1,2}=t_{\rm L}\mp i t_{\rm R}$, where \(t_{\rm L}\) and \(t_{\rm R}\) denote the couplings to the Majorana modes on the left and right chains, respectively (see
\cref{fig:Schematics}). These matrix elements are controlled by the magnetic orientation \((\upzeta,\theta)\) of the control adatom and take the form
\begin{equation}
t_{\rm L(R)}
=
e^{i\upzeta/2}
\cos\left(\frac{\theta}{2}\right)
F_{\rm L(R)}
+
e^{-i\upzeta/2}
\sin\left(\frac{\theta}{2}\right)
G_{\rm L(R)}\,,
\label{eq:tL_tR}
\end{equation}
where \(F_{\rm L(R)}\) and \(G_{\rm L(R)}\) encode the microscopic properties of the substrate and the corresponding Majorana wave functions~\cite{Awoga2024}. Since the coefficients \(F_{\rm L(R)}\) and \(G_{\rm L(R)}\) inherit the Majorana wave-function amplitude at the position of the control adatom, the Shiba--Majorana couplings $|t_{\rm L,R}|$ decay asymptotically in a similar (exponential) oscillatory form of the direct tip--Majorana coupling \(t_{\rm M}\), since  both are governed by the localized Majorana wave function, as shown in \cref{eq:tM_tS_scaling}. The magnetic orientation \((\upzeta,\theta)\) controls the relative magnitude and phase of \(t_{\rm L,R}\), whereas the device geometry, through the distance \(R\), determines their overall strength.

The effective Hamiltonian in \cref{eq:HamiltonianLowEenergy} retains only the control-Shiba state, the two inner Majorana modes, and their coupling to the superconducting tip. Its validity therefore requires the couplings to the Majorana sector to remain perturbative with respect to the topological minigap,
\begin{equation}
|t_{\rm M}|,\;|t_{\rm L,R}|
\ll
\Delta_{\rm top}\,,
\end{equation}
such that virtual transitions to the higher-energy Shiba-band states can be neglected. Moreover, all retained low-energy scales are assumed to be small compared with the parent superconducting gap \(\Delta\), ensuring that the quasiparticle continuum contributes only a smooth background to the supercurrent, while the nonanalytic features originate exclusively from the discrete Shiba--Majorana spectrum.

Diagonalizing \(H_{\rm low}\) yields the hybridized in-gap energies \(E_{\rm M}\) and \(E_{\rm S}\), which evolve from the bare energies \(\epsilon_{\rm M}\) and \(\epsilon_{\rm S}\) as the couplings \(t_{1,2}\) are varied; see \cref{eq:AppendixPoles}. Coupling the system to the superconducting tip further modifies these levels and makes their energies dependent on the superconducting phase difference. This phase dependence generates an equilibrium supercurrent between the tip and the substrate.
\begin{figure*}[t]
    \centering
    \includegraphics[width=0.95\linewidth,height=0.27\textwidth]{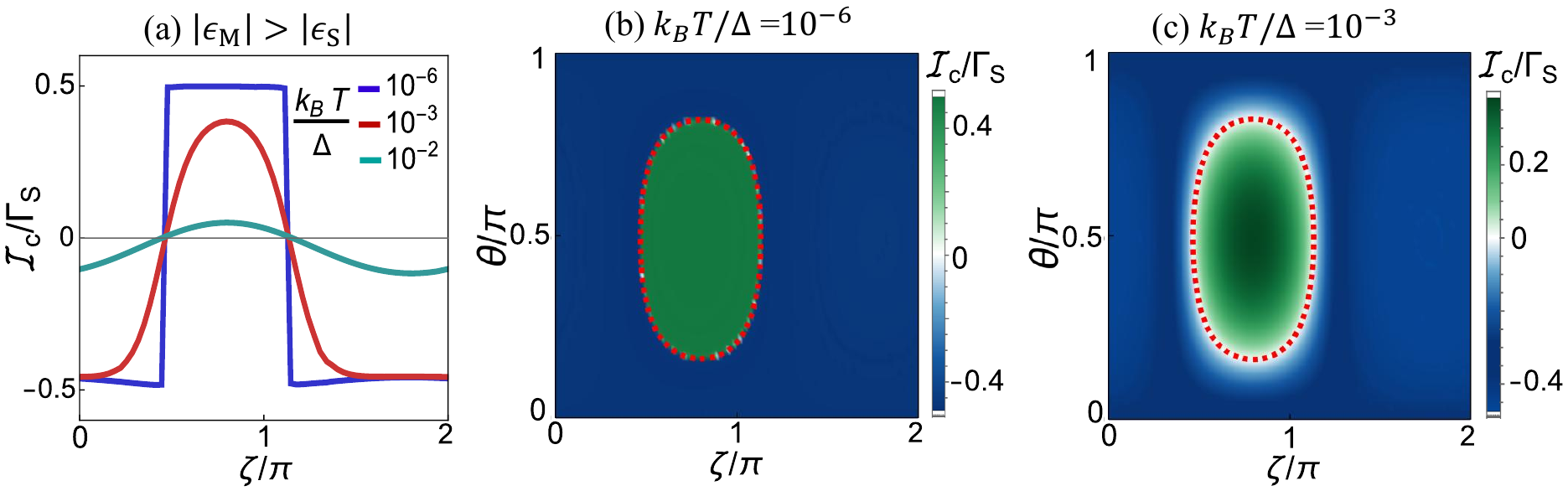}
    \caption{Effect of finite temperature and adatom orientation on the Shiba-mediated Josephson current at weak tunneling regime. (a) Critical supercurrent with respect to the adatom magnetization angle $\upzeta$ at $\theta=\pi/2$ for different temperatures showing the signature of the quantum phase transition, which smears out as $T$ increases. We consider the electron tunneling rate from the tip--to--Shiba state $\Gamma_{\mathrm{S}}/\Delta=10^{-4}$ at tip--to--Majorana tunneling rate $\Gamma_{\mathrm{M}}/\Delta=0$. The remaining parameters are the same as in \cref{fig:2}(b). (b) 2D plot of critical supercurrent for the variation of the magnetization angles of the adatom $(\upzeta, \theta)$ at $k_B T/\Delta=10^{-6}$, at the same parameters as in (a). The sign change in the current represents the topological quantum phase transition, followed by the sign change of the renormalized Majorana energy $E_{\mathrm{M}}$ along the critical $\theta_{\mathrm{c}}$ and $\upzeta_{\mathrm{c}}$, as indicated by the red dashed line. (c) Same as in (b), but at $k_B T/\Delta=10^{-3}$ that shows the thermal smearing effect through the broaden close-to-zero supercurrent region around the topological phase transition.}
    \label{fig:3}
\end{figure*}
Upon integrating out the superconducting tip, it is convenient to introduce the tunneling scales
\begin{equation}
\Gamma_{\rm S}
=
\pi\nu_{0}t_{\rm S}^{2},
\qquad
\Gamma_{\rm M}
=
\pi\nu_{0}t_{\rm M}^{2},
\label{eq:Gamma_SM}
\end{equation}
where \(\nu_{0}\) is the normal-state density of states of the tip. These quantities characterize the strength of the coherent superconducting self-energy induced in the Shiba and Majorana sectors.
We first consider the regime in which the tip--Majorana coupling is negligible as $\Gamma_{\rm M}= 0$ and the tip couples predominantly to the control-Shiba state with finite $\Gamma_{\rm S}$. After expanding the low-energy eigenvalues in the weak tunneling limit (see Appendix~\ref{Appendix:CurrentDerivation}), we obtain the supercurrent at zero temperature
\begin{equation}
\mathcal I(\phi)
=
\Gamma_{\rm S}
\frac{
\epsilon_{\rm S}
+
\epsilon_{\rm M}\,
{\rm sgn}(E_{\rm M})
}{
2\left(
E_{\rm S}+|E_{\rm M}|
\right)
}
\sin\phi\,.
\label{eq:SupercurrentExpressionZeroTSystem}
\end{equation} 
In this case, the current--phase relation is sinusoidal, while higher harmonics may arise in the strong tunneling limit. 

This weak-coupling supercurrent at lowest order of $\Gamma_{\rm S}$ directly relates the supercurrent to the parity-changing zero-energy crossings of the Shiba--Majorana spectrum. When \(E_{\rm M}\) changes sign, the ground-state occupation changes and the zero-temperature current develops a discontinuity. For $|\epsilon_{\rm M}|<|\epsilon_{\rm S}|$, this produces an abrupt change in the magnitude of the supercurrent, as illustrated in the left inset of \cref{fig:Schematics}. By contrast, for $|\epsilon_{\rm M}|>|\epsilon_{\rm S}|$, the state crossing zero has a stronger control-Shiba character and is therefore more directly coupled to the tip, resulting in a characteristic reversal of the supercurrent; see the right inset of \cref{fig:Schematics}.

To describe arbitrary tip coupling and finite temperature, we evaluate the equilibrium current using the Keldysh Green-function formalism~\cite{Jauho1994,Sun2000a,Kamenev2005,Villas2020,Chakraborty2023,Cheng2023,Debnath2024,Debnath2025}. The current flowing out of the superconducting tip is defined by 
$\mathcal I=
-\left\langle
\frac{d\mathcal N_{\rm tip}}{dt}
\right\rangle
=
i\left\langle
\left[
\mathcal N_{\rm tip},
H_{\rm tip-low}
\right]
\right\rangle$,
where
$\mathcal N_{\rm tip}
=
\sum_{\mathbf{k},\sigma}
a_{\mathbf{k}\sigma}^{\dagger}
a_{\mathbf{k}\sigma}$ is the electron-number operator of the tip. Expressing the supercurrent in terms of the Nambu-space lesser Green's function and using the equilibrium fluctuation--dissipation relation together with Dyson's equation, yields the full current--phase relation. Details of this calculation are provided in Appendix~\ref{Appendix:KeldyshMethod}. Throughout the manuscript, all energies are expressed in units of the superconducting gap \(\Delta\).

\section{Results on Quantum Phase Transitions} \label{Results}
In this section, we investigate the supercurrent characteristics using the Keldysh Green's function formalism, beyond the low-tunneling regime $(\Gamma/\Delta>10^{-4})$ for different orders of the tunneling rate $\Gamma_{\mathrm{S}}/\Delta$ at $\Gamma_{\mathrm{M}}/\Delta=0$. To observe the effect of the sign change of $E_\mathrm{M}$ associated with the quantum phase transition, we depict the supercurrent with respect to the adatom azimuthal angle $\upzeta$ in \cref{fig:2}(a) and \cref{fig:2}(b) for two energy regimes of Majorana $(\epsilon_\mathrm{M})$ and Shiba $(\epsilon_\mathrm{S})$ states, for $\abs{\epsilon_{\text{M}}}<\abs{\epsilon_{\text{S}}}$ and $\abs{\epsilon_{\text{M}}}>\abs{\epsilon_{\text{S}}}$, respectively. Concretely, we demonstrate discontinuities in the supercurrent while tuning the control adatom's magnetic orientation $\upzeta$ for the scenarios $\abs{\epsilon_{\text{M}}}\gtrless\abs{\epsilon_{\text{S}}}$, reflecting the changes in the ground state parity of the system. Interestingly, the supercurrent jump occurs at specific azimuthal angles $\upzeta_{c1}$ and $\upzeta_{c2}$, at a fixed polar angle $\theta$, which we consider here as $\pi/2$. These quantum phase transition points, $\upzeta_{c1}$ and $\upzeta_{c2}$, for the supercurrent jump are independent of the tunneling rate $\Gamma_{\mathrm{S}}/\Delta$, but sensitive to all other energy scales. As the tunneling rates $\Gamma_\mathrm{S(M)}/\Delta$ are functions of the tip--Shiba $(t_{\rm S})$ and tip--Majorana $(t_{\rm M})$ couplings, controlling the tip distance to the control adatom modifies $\Gamma_\mathrm{S(M)}/\Delta$. For a larger transmission rate, the hybridized in-gap energies are modified, leading to zero energy crossing of $E_{\rm M}$. As a consequence, we obtain sign changes in the supercurrent at $\Gamma_\mathrm{S}/\Delta=10^{-2}$, see \cref{fig:2}(c) and \cref{fig:2}(d). It is important to mention that we consider the equal particle-hole weight of the Shiba state in the tunneling Hamiltonian in \cref{eq:HamiltonianLowEenergy}, i.e., $u=h$. Particle-hole asymmetry significantly modifies the supercurrent quantitatively, yet maintains the signature of the quantum phase transition being a jump in supercurrent, see Appendix~\ref{Appendix:Particle-hole asymmetry}.
\begin{figure*}
    \centering
    \includegraphics[width=\linewidth,height=0.27\textwidth]{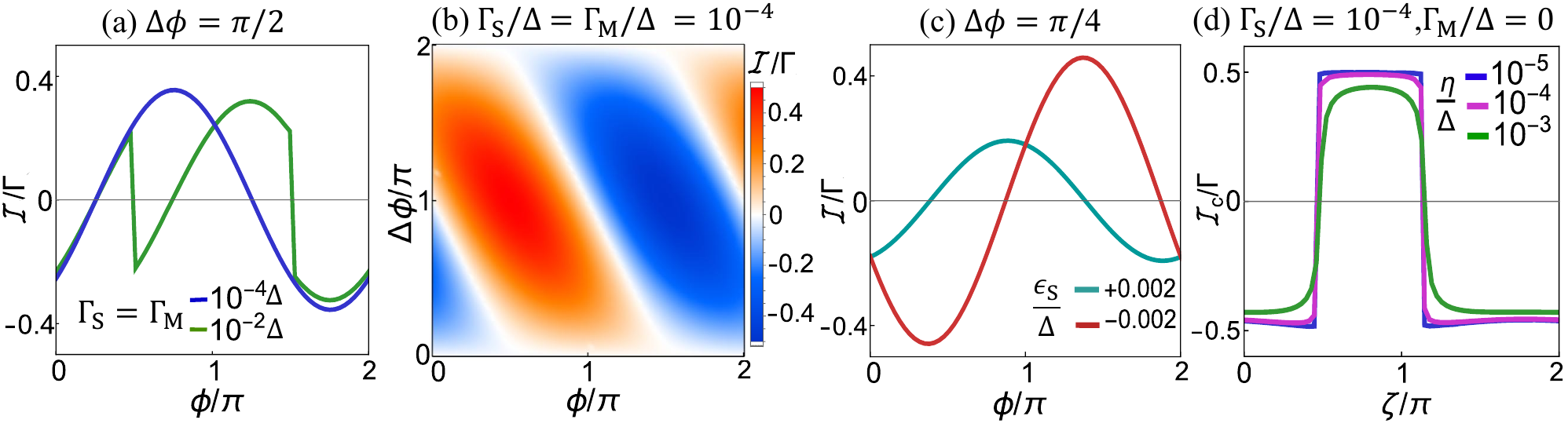}
    \caption{Josephson current in the presence of both tip--Shiba and tip--Majorana coupling.
    (a) Supercurrent as a function of the phase $\phi$ entering the tip--Shiba tunneling, for several total couplings $\Gamma=\Gamma_{\mathrm S}+\Gamma_{\mathrm M}$, (b) Supercurrent map in the $(\phi,\Delta\phi)$ plane for $\Gamma/\Delta=10^{-4}$. Here, $\phi$ and $\phi+\Delta\phi$ denote the phases associated with the two effective tunneling paths from the tip to the Shiba and Majorana sectors. For panels (a) and (b), we use $\epsilon_{\mathrm M}=-0.2\Delta$, $\epsilon_{\mathrm S}=0.002\Delta$, and $(\upzeta,\theta)=(\pi/2,\pi/2)$. (c) Supercurrent for positive and negative Shiba energies, showing the supercurrent changes with the sign of $\epsilon_{\mathrm S}$. The parameters are $\Gamma/\Delta=10^{-4}$, $\epsilon_{\mathrm M}=-0.2\Delta$, and $(\upzeta,\theta)=(\pi/2,\pi/2)$. (d) Critical current as a function of the adatom magnetization angle $\upzeta$ at $\theta=\pi/2$, for different values of the broadening parameter $\eta$. The calculation is performed at low temperature, $k_{\mathrm B}T/\Delta=10^{-5}$, at $\Gamma_{\mathrm S}/\Delta=10^{-4}$ and $\Gamma_{\mathrm M}/\Delta=0$. The remaining parameters are the same as in Fig.~\ref{fig:2}(b).}
    \label{fig:4}
\end{figure*}

Next, we investigate the influence of a finite temperature $T$ on the topological phase transition, treating temperature as a control parameter that suppresses the STM current through the Shiba state and ultimately drives it to zero at sufficiently high $T$. This behavior originates from thermal smearing of the Fermi-Dirac function, which reduces the occupation imbalance needed for subgap transport and thereby weakens the supercurrent jump, causing the quantum phase transition signature to gradually fade, see \cref{fig:3}(a). As our setup is controlled by the orientation of the magnetic adatom due to Shiba--Majorana hybridization via $t_{\text{L}}(\upzeta, \theta)$ and $t_{\text{R}}(\upzeta, \theta)$ (see \cref{eq:tL_tR}), the variation of the supercurrent in the parameter space $(\upzeta, \theta)$ at different temperatures provides a phase diagram of the critical transition points, as shown in \cref{fig:3}(b) and \cref{fig:3}(c). The appearance of an extended region with close-to-zero supercurrent at $k_{B}T/\Delta=10^{-3}$ signifies the smearing effect of the supercurrent jump, as observed in \cref{fig:3}(a). In the weak-coupling regime, the quantum phase transition detected by a jump in the supercurrent in the parameter space $(\upzeta, \theta)$, agrees with the topological phase transition characteristics of a Majorana--Shiba hybridized state presented in Ref.~\cite{Awoga2024}. 
\section{Effects of tip--Majorana coupling and Dissipation}\label{TipMajoranaCoupling}
As considered in the Hamiltonian $H_{\mathrm{tip-low}}$ in \cref{eq:HamiltonianLowEenergy}, the superconducting tip can also be directly coupled to the Majorana modes. In this section, we allow the tunneling rate $\Gamma_\mathrm{M}/\Delta$ to be finite and the supercurrent-driving phase $\phi_2=\phi+\Delta\phi$ of the tip connected to the Majorana modes to vary and study their effects on the supercurrent numerically. Since the superconducting phases $\phi$ and $\phi+\Delta\phi$ drive the supercurrent from the tip to the Shiba and Majorana states with tunneling rates $\Gamma_{\rm S}/\Delta$ and $\Gamma_{\rm M}/\Delta$, respectively, for finite supercurrent tunneling to both states we define the total tunneling rate $\Gamma/\Delta=(\Gamma_{\rm S}+\Gamma_{\rm M})/\Delta$. As the additional phase $\Delta\phi$ induced to the tip-Majorana tunneling is unknown, we first fix $\Delta\phi=\pi/2$ and study the variation of supercurrent with superconducting phase $\phi$ at different $\Gamma/\Delta$ in \cref{fig:4}(a). We obtain a finite supercurrent at $\phi=0$, which signifies that in the presence of additional phase in the tip--Majorana tunneling, the supercurrent no longer responds to the single superconducting phase $\phi$, but to an effective phase set by the vector sum of the two pairing amplitudes, leading to a coupling-dependent shift of the phases. We obtain a sharp sign change in supercurrent for $\Gamma/\Delta=10^{-2}$ associated with zero-energy crossings as also observed in \cref{fig:2}(c) for $\Gamma_{\rm M}/\Delta=0$.

To investigate the effect of the additional $\Delta\phi$ phase on the total supercurrent, we plot the supercurrent for the simultaneous variation of $\phi$ and $\Delta\phi$ in \cref{fig:4}(b), considering $\Gamma/\Delta=10^{-4}$. We find that the spectrum exhibits zero supercurrent nodal lines, indicating that the Josephson response is governed by interference between two tunneling phases. As the sign of supercurrent for the tip--Majorana coupling (see \cref{fig:6}(a)) is opposite to the supercurrent for tip--Shiba tunneling (see \cref{fig:2}(c)), in the weak tunneling regime, i.e., $\Gamma/\Delta=10^{-4}$, we interestingly observe that the supercurrent follows
\begin{align}
    \mathcal{I}/\Gamma =
    \mathrm{sin}\left(\frac{\Delta\phi}{2}\right)\mathrm{cos}\left(\phi+\frac{\Delta\phi}{2}\right)/4,
   \label{eq:curr_delphi}
\end{align}
matching the result obtained by the Keldysh Green's function formalism in \cref{fig:4}(b). We also study the supercurrent for a stronger tunneling rate and different adatom rotational angles, presented in Appendix~\ref{Appendix:two phases}. In \cref{fig:4}(c), we choose $\Delta\phi=\pi/4$ to study the effect of the sign of the bare Shiba energy on the total supercurrent at $\Gamma/\Delta=10^{-4}$. Although the supercurrent exhibits a $0$--$\pi$ transition~\cite{Kulik1965, Baselmans1999, vanDam2006, Delagrange2016} at $\Gamma_{\rm M}/\Delta=0$ (see \cref{fig:5}(b)), a finite $\Delta\phi$ shifts the current-phase relation at $\Gamma_{\rm S}/\Delta=\Gamma_{\rm M}/\Delta=10^{-4}$.

It is important to note that the sharp jump predicted by the ideal equilibrium calculation is expected to be broadened in realistic experiments~\cite{Becker2006,Jdira2008,Steiner2023}. Microscopically, this broadening reflects dissipative processes that give the relevant subgap state a finite lifetime, such as quasiparticle poisoning or other inelastic/environmental relaxation channels. Within the Green's function framework, these effects can be incorporated phenomenologically by introducing a finite broadening parameter \(\eta\), as specified in \cref{eq:g_MM}, and the supercurrent can then be evaluated following Ref.~\cite{Shen2024}. Therefore, the parameter \(\eta\) should be understood as an effective measure of dissipation, regardless of its microscopic origin. As \(\eta\) increases, phase coherence is reduced, which suppresses the coherent supercurrent and smears the sharp jump associated with the quantum phase transition, as shown in \cref{fig:4}(d).

\section{Implementation and parity discrimination}\label{Implementation}
Having established the phase-dependent Shiba spectrum and transport properties, we next address how the setup can be implemented without direct control of the superconducting phase difference, which is typically unavailable in realistic STM experiments. Instead, the superconducting junction is generally operated in a current-biased configuration. To bridge the gap between the theoretical framework for phase bias developed above and the experimentally accessible observables, we consider the junction connected in series with a large external resistance ($R_{\rm ext} \approx 10^{9}\Omega$) and driven by a voltage bias $V_{\rm b}$. In this high-impedance limit, the current is approximately fixed $\mathcal{I} \approx V_{\rm b}/R_{\rm ext}$ and remains largely insensitive to the microscopic dynamics of the junction. The transition from a phase-biased description to the current-biased regime is established via an effective macroscopic Hamiltonian for the phase coordinate~\cite{ScienceMartinis88}
\begin{equation}
\mathcal{H}_{\rm eff} = \frac{\hat{Q}^2}{2C} + E_{\mathrm{GS}}(\hat{\phi}, \mathcal{P}) - \frac{\mathcal{I}}{2}\hat{\phi}\,,
\end{equation}
where $E_{\mathrm{GS}}(\hat{\phi}, \mathcal{P})$ is the parity-dependent ground state energy given by Eq.~(\ref{eq:SupercurrentExpressionZeroT}), and $C$ is the junction capacitance. The variable $Q$ denotes the macroscopic charge accumulated across the junction, acting as the canonical conjugate momentum to the phase operator ($[\hat{\phi}, \hat{Q}] = 2\,i$). In the adiabatic limit, $\phi$ acts as a coordinate trapped in the local minima of the washboard potential $U(\phi,\mathcal{P}) = E_{GS}(\phi, \mathcal{P}) - \frac{\mathcal{I}}{2}\phi$. The junction sustains a zero-voltage state as long as the phase remains trapped in a local minimum. A measurable DC voltage $V =\langle \partial_t\hat{\phi} \rangle/2$ emerges only when the bias current drives the system out of the stationary regime.

A fundamental detection challenge for detecting the ground state parity arises if the Shiba-mediated current is the sole contribution to transport. Because the two parity sectors differ primarily by a $\pi$-phase shift in their energy landscapes, i.e., $E_{\rm{GS}}(\phi, \mathcal{P}=1) = E_{\rm{GS}}(\phi + \pi, \mathcal{P}=-1)$, the maximum current-sustaining capacity and thus the measured switching currents $\mathcal{I}_{\rm sw}$ (at which the junction transitions from a zero-voltage to a finite-voltage state), would be identical. In such a symmetric scenario, a parity switch would produce only a transient voltage pulse rather than a distinguishable steady DC signal.
However, in a realistic STM geometry, the localized Shiba state is embedded in a junction where the bulk superconducting tips also contribute to the total supercurrent. This bulk channel provides a global phase reference $E_{\rm bulk}(\phi) = -E_{J} \cos \phi$, which breaks the symmetry between the parity sectors. The total effective potential becomes
\begin{equation}
U_{\rm tot}(\phi) = -E_J \cos \phi + E_{\rm{GS}}(\phi, \mathcal{P}) - \frac{\mathcal{I}}{2}\phi\,.
\end{equation}
Because $E_{\rm{GS}}(\phi, \mathcal{P})$ is generally a non-sinusoidal function of $\phi$, the addition of the bulk term ensures that the potential landscape explicitly depends on $\mathcal{P}$. Specifically, the critical current $\mathcal{I}_{\rm c}(\mathcal{P})$ at which the local minima disappear reflects constructive or destructive interference between the two channels.
\begin{equation}
\mathcal{I}_{\rm c}(\mathcal{P}) = \max_\phi \left| 2\frac{\partial}{\partial \phi} \left (-E_J \cos \phi + E_{\rm{G}S}(\phi, \mathcal{P}) \right) \right|\,.
\end{equation}
Since generically $\mathcal{I}_{\rm c}(\mathcal{P}=1) \neq \mathcal{I}_{\rm c} (\mathcal{P}=-1)$, the two parity sectors have different breakdown thresholds. Consequently, the transition between the parity sectors manifests itself as a discontinuous jump in the measured switching current, allowing for direct experimental discrimination of the $0-\pi$ transition.

Let us emphasize that at finite temperature, the observable switching current $\mathcal{I}_{\rm sw}$ is a stochastic quantity that typically remains lower than the deterministic critical current ($\mathcal{I}_{\rm sw} < \mathcal{I}_
{\rm c}$) due to thermal activation over parity-dependent barriers $\Delta U(\mathcal{I}, \mathcal{P})$. Furthermore, in the extreme small-capacitance limit characteristic of STM geometries (${\sim}10^{-15}$--$10^{-18}$ F), large phase fluctuations dominate. In this regime, a comprehensive quantitative treatment of the dissipative transport requires more elaborate approaches like the \(P(E)\)-theory description of energy exchange with the high-impedance electromagnetic environment~\cite{Karan2022}. The environment converts phase fluctuations into inelastic Cooper-pair tunneling events, producing a finite dc current and a measurable conductance whose thresholds and weights reflect the parity-dependent many-body spectrum of the junction. Thus, the \(P(E)\) description can provide a direct parity-sensitive readout channel. At the same time, since the Josephson coupling enters only perturbatively in this description, it does not act as a coherent control knob for tuning or hybridizing the low-energy states. The role of this regime is therefore primarily spectroscopic: it probes the existing parity-dependent spectrum rather than engineering it. 

Although a full treatment of these quantum fluctuations and the possible crossover into a phase-diffusion regime is beyond the scope of this work, the qualitative mechanism for parity discrimination remains robust. The bulk supercurrent provides the necessary symmetry-breaking reference that maps the parity-dependent ground-state energy \(E_{\rm GS}(\phi,\mathcal P)\) onto distinct critical-current or conductance signatures.
  
\section{Conclusions} \label{Conclusion}
Our results establish that a controllable magnetic adatom with Shiba–Majorana hybridization provides a robust mechanism for detecting topological quantum phase transitions via discontinuities and sign changes in the supercurrent, highlighting a viable route for supercurrent-based manipulation in qubit architectures. Here, the quantum phase transition is highly sensitive to the microscopic system parameters, and the orientation of the magnetization of the adatom serves as a control knob for probing the phase transition of the Shiba--Majorana state through the supercurrent in the Josephson junction. We investigate the STM probed supercurrent using Dyson’s equation of motion combined with the fluctuation-dissipation theorem within the Keldysh Green’s function formalism, for different tunneling regimes and adatom magnetization angles. We verified the consistency of the weak tunneling rate limit results through an analytical low-energy theory that matches exactly with the numerical Keldysh results. Finite temperatures play a key role by broadening the sharp jump of the supercurrent at the critical magnetization angle of the Shiba state, ultimately leading to a vanishing supercurrent at higher temperatures. We further demonstrate that tuning the bare energy of the Shiba state enables a controllable $0-\pi$ phase transition in the supercurrent, which gets modified in the presence of the effective additional phase to the tip-Majorana coupling. We also show that the effect of this direct coupling of the superconducting tip to the Majorana modes does not alter the quantum phase transition characteristics of the supercurrent.

Our study provides a route to experimental implementation of current- and phase-biased superconducting STM tips as a probe for detecting topological phase transitions in adatom architectures. This may support future Majorana braiding protocols~\cite{Burrello2013, Souto2022} via controlled manipulation of supercurrent through adatom rotation. Beyond detection, our model also offers a promising platform for realizing a topological Josephson diode~\cite{Kotetes2026, Lu2023}. The nonreciprocity of the supercurrent may naturally emerge from the hybridization between Shiba and Majorana states, in conjunction with time-reversal symmetry breaking induced by the magnetic adatom. With the inclusion of additional symmetry-breaking ingredients, this setup holds the possibility to enable diode functionality, positioning it as a building block for quantum micro-fabricated, temperature-sensitive superconducting devices in which magnetism, topology, and superconductivity are intertwined.

\section{Acknowledgments}
D.\ D.\ acknowledges funding by the NGP network on spin, topology and superconductivity, the APS-EPS-FECS-ICTP Travel Award Fellowship Programme (ATAP), Trieste, Italy, and Deutsche Forschungsgemeinschaft (DFG, German Research Foundation) under Germany’s Excellence Strategy –EXC-2123/2 QuantumFrontiers – 390837967.
D.\ D.\ and P.\ D.\ acknowledge the Department of Space, Government of India for all support at PRL. 
I. I. and T. P. acknowledge funding by the Cluster of Excellence
`Advanced Imaging of Matter' (EXC 2056, project ID 390715994) of the Deutsche Forschungsgemeinschaft (DFG). 
T. P. acknowledges funding from the European Union (ERC Starting Grant QUANTWIST, project number 101039098). 
M. T.~acknowledges support from the National Science Center (Poland) OPUS Grant No. 2021/42/B/ST3/04475, and the Foundation for Polish Science project ``MagTop'' (No. FENG.02.01-IP.05-0028/23) cofinanced by the European Union from the funds of Priority 2 of the European Funds for a Smart Economy Program 2021-2027 (FENG), and by the NAWA Bekker Grant No. BPN/BEK/2024/1/00310 (Poland).
\appendix
\renewcommand{\thefigure}{A\arabic{figure}}

\section{Derivation of supercurrent for Shiba-Majorana hybridized state}\label{Appendix:CurrentforShiba}

In this Appendix, we formulate the Josephson current through a superconducting
tip coupled locally to a magnetic adatom. We first derive the full retarded
Green's function of the tip-dressed Shiba state and use it to compute the
equilibrium current, including both the subgap Shiba-pole contribution and
the quasiparticle-continuum contribution. We then explain how the same
microscopic description can be projected onto a low-energy effective model,
where the tip couples directly to the Shiba quasiparticle with an effective
tunneling amplitude \(t_{\rm S}\).

\subsection{Retarded Green's function and full Josephson current}

We now describe the nonperturbative calculation of the Josephson current
through a superconducting tip placed directly above the magnetic adatom. The
tip is treated as a superconducting reservoir with phase difference \(\phi\)
with respect to the substrate. Throughout this section, we focus on the local
Green's function at the impurity position, since the tunneling Hamiltonian
couples the tip only to this local degree of freedom.

For a classical magnetic impurity in an \(s\)-wave superconductor, the local
retarded Green's function in the relevant Shiba block can be written as ($\hbar=1$) \cite{RubyPRL15}
\begin{equation}
g_S^R(\omega)
=
\pi\nu_0
\frac{
(z+\alpha s_R)\tau_0+\Delta\tau_x
}{
2\alpha z-(1-\alpha^2)s_R
}\,,
\end{equation}
where $z=\omega+i0^+$ and $s_R(\omega)=\sqrt{\Delta^2-z^2}$. Here, 
 the branch is chosen so that \(\mathrm{Re}\,s_R>0\) is inside the gap,  \(\nu_0\) is the density of the normal-state of the states, \(\alpha=\pi\nu_0JS\),
and \(\tau_i\) are Pauli matrices in the Nambu space. In the absence of the tip,
the pole of \(g_S^R\) gives the bare Shiba energy
\begin{equation}
\epsilon_{\rm S}
=
\Delta\frac{1-\alpha^2}{1+\alpha^2}\,.
\label{Shiba_energy}
\end{equation}
The superconducting tip contributes to the local self-energy
\[
\Sigma_{\rm tip}^R(\omega,\phi)
=
|t_{\rm tip}|^2 g_{\rm tip}^R(\omega,\phi)\,,
\]
where
\begin{equation}
g_{\rm tip}^R(\omega,\phi)
=
-\pi\nu_0
\frac{
z \tau_0+
\Delta\left(\cos\phi\,\tau_x-\sin\phi\,\tau_y\right)
}{
s_R(\omega)
}\,.
\end{equation}
The full retarded Green's function at the impurity position is therefore
\begin{equation}
G^R(\omega,\phi)
=
\left[
(g_S^R(\omega))^{-1}-\Sigma_{\rm tip}^R(\omega,\phi)
\right]^{-1}\,,
\end{equation}
so that the poles of the dressed Shiba state are obtained from
\begin{equation}
D_R(\omega,\phi)
=\det[G^R(\omega,\phi)]^{-1}=0\,,
\end{equation}
or explicitly
\[
\left[(1+\gamma_{\rm tip})\omega+\alpha\sqrt{\Delta^2-\omega^2}\right]^2
=
\Delta^2
\left(
1-2\gamma_{\rm tip}\cos\phi+\gamma_{\rm tip}^2
\right)\,,
\]
for $\qquad |\omega|<\Delta$, where $\gamma_{\rm tip}=(\pi\nu_0|t_{\rm tip}|)^2$. The branch continuously connected to the bare Shiba state at
\(\gamma_{\rm tip}=0\) is
\begin{equation}
E_{\rm S}(\phi)
=
\Delta
\frac{
(1+\gamma_{\rm tip})B(\phi)
-\alpha C(\phi)
}{
\alpha^2+(1+\gamma_{\rm tip})^2
}\,,
\end{equation}
where
\begin{align}
B(\phi)=\sqrt{1-2\gamma_{\rm tip}\cos\phi+\gamma_{\rm tip}^2}\,,\nonumber\\
C(\phi)=
\sqrt{
\alpha^2+(1+\gamma_{\rm tip})^2-B^2(\phi)
}\,.
\end{align}
At \(\gamma_{\rm tip}=0\), this expression reduces to the bare Shiba energy $\epsilon_{\rm S}$ in \cref{Shiba_energy}.

The full equilibrium current can be obtained from the phase derivative of
the fermionic ground-state energy. In the real-frequency formulation this
can be written in terms of the retarded Green's function  \cite{AltlandSimons2010}
\begin{equation}
\mathcal{I}_{\rm full}(\phi)
=
-e
\int_{-\infty}^{0}
\frac{d\omega}{\pi}\,
\partial_\phi
\mathrm{Im}\,
\ln
\det
[G^R(\omega,\phi)]^{-1}\,.
\end{equation}
Since the phase dependence enters only through the tip self-energy, this
can also be written as
\[
\mathcal{I}_{\rm full}(\phi)
=
e
\int_{-\infty}^{0}
\frac{d\omega}{\pi}\,
\mathrm{Im}\,
\mathrm{Tr}
\left[
G^R(\omega,\phi)
\partial_\phi \Sigma_{\rm tip}^R(\omega,\phi)
\right]\,,
\]
up to the overall sign convention for the direction of positive current.
This expression contains both the discrete Shiba pole contribution and the
continuum contribution from the superconducting quasiparticle branch cuts.  
The total current may be written as
\[
\mathcal{I}_{\rm full}(\phi)
=
\mathcal{I}_{\rm Shiba}(\phi)+\mathcal{I}_{\rm cont}(\phi)\,.
\]
The first term comes from the occupied subgap pole. In the reduced
\(2\times2\) Nambu-block convention used above, the zero-temperature pole
contribution is
\begin{equation}
\mathcal{I}_{\rm Shiba}(\phi)
=
-\frac{e}{2}
\mathrm{sgn}\!\left[E_{\rm S}(\phi)\right]
\partial_\phi E_{\rm S}(\phi)\,.
\end{equation}
The factor \(1/2\) accounts for the Nambu double counting in the local
determinant formulation. 

Using the explicit expression for \(E_{\rm S}(\phi)\), we find that at a parity-changing point $E_{\rm S}(\phi)=0$, the zero-temperature current has a discontinuity. Equivalently, the crossing condition is
\[
\alpha^2=
1-2\gamma_{\rm tip}\cos\phi+\gamma_{\rm tip}^2\,.
\]

\begin{figure}
    \centering
    \includegraphics[width=\linewidth,height=0.45\linewidth]{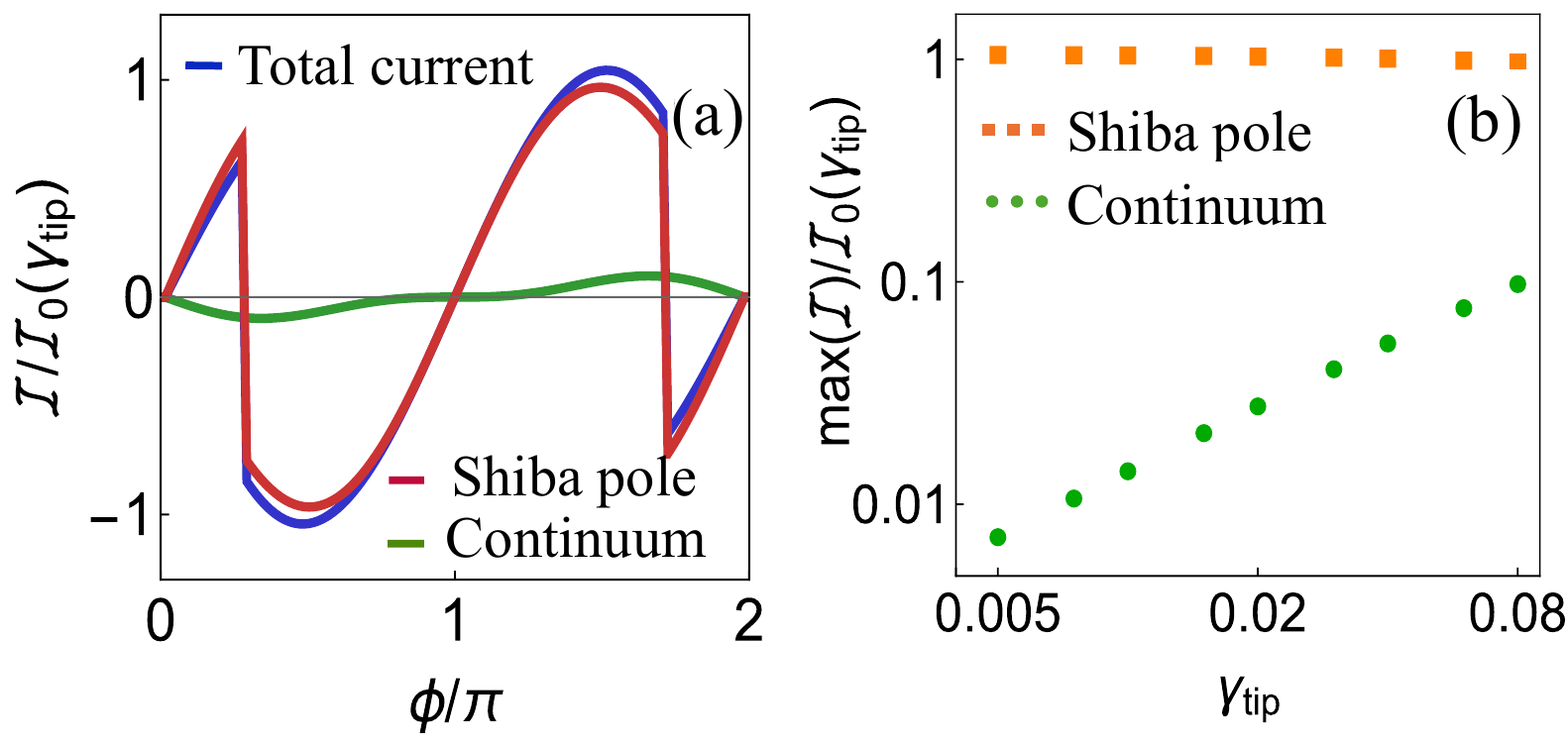}
   \caption{Supercurrent through a single Shiba state coupled to a superconducting tip.(a) Phase dependence of the total current, the Shiba-pole contribution, and the continuum contribution, normalized by \(\mathcal{I}_0(\gamma_{\rm tip})=\gamma_{\rm tip}\Delta/2\). The discontinuities in the Shiba-pole contribution occur when the dressed Shiba level crosses zero energy, corresponding to a change in the occupied subgap branch. The full current differs from the Shiba-pole contribution by a smooth continuum background. The parameters are \(\alpha=0.95\) and \(\gamma_{\rm tip}=0.01\). (b) Maximum absolute value of the Shiba-pole and continuum contributions as a function of \(\gamma_{\rm tip}\), normalized by the same scale \(\mathcal{I}_0(\gamma_{\rm tip})\). The Shiba-pole contribution remains linear in \(\gamma_{\rm tip}\) at weak coupling, whereas the continuum contribution is subleading and scales as \(\mathcal{I}_{\rm cont}\propto\gamma_{\rm tip}^2\) for small \(\gamma_{\rm tip}\). Consequently, \(\mathcal{I}_{\rm cont}/\mathcal{I}_0\propto\gamma_{\rm tip}\) in the normalized plot.}
\label{fig:5}
\end{figure}

The continuum contribution is defined as the remainder
\[
\mathcal{I}_{\rm cont}(\phi)
=
\mathcal{I}_{\rm full}(\phi)-\mathcal{I}_{\rm Shiba}(\phi)\,.
\]
Equivalently, it can be written directly as the branch-cut contribution
outside the superconducting gap,
\[
\mathcal{I}_{\rm cont}(\phi)
=
-e
\int_{-\infty}^{-\Delta}
\frac{d\omega}{\pi}\,
\partial_\phi
\mathrm{Im}\,
\ln
\det[G^R(\omega,\phi)]^{-1}\,,
\]
again up to the same current-direction convention. This term represents the phase-dependent redistribution of continuum
quasiparticle states induced by the coupling to the superconducting tip. In
\cref{fig:5}, we evaluate this contribution
numerically and compare it with the Shiba-pole contribution. In the
weak-tunneling regime, the continuum part is subleading, while the leading
current is controlled by the dressed Shiba pole.

In the weak-tunneling limit, expanding the dressed Shiba pole to the leading
order in \(\gamma_{\rm tip}\) gives 
\begin{equation}
E_{\rm S}(\phi)
=
\epsilon_{\rm S}
-
\frac{2\gamma_{\rm tip}\Delta}{1+\alpha^2}\left[\cos\phi
-
\frac{1-\alpha^2}
{1+\alpha^2}\right]
+
O(\gamma_{\rm tip}^2)\,.
\end{equation}
Close to the bare Shiba crossing, \(\alpha\simeq1\), this reduces to
\[
E_{\rm S}(\phi)
\simeq
\Delta(1-\alpha)
-
\gamma_{\rm tip}\Delta\cos\phi\,.
\]
Thus, the tip-induced phase-dependent shift is of order
$\delta E_{\rm S}\sim \gamma_{\rm tip}\Delta$. Since \(\gamma_{\rm tip}\) is dimensionless in the present normalization,
the weak tunneling condition near a zero-energy bare Shiba state is simply
\[
\gamma_{\rm tip} = \left(\pi \nu_0 |t_{\rm tip}|\right)^2  \ll1\,,
\]
which, interestingly, does not involve the SC gap \(\Delta\). Instead, the gap
only sets the overall energy scale of the Shiba pole and of its
tip-induced shift. The corresponding leading Shiba current is
\begin{equation}
\mathcal{I}_{\rm Shiba}(\phi)
\simeq
-e\Delta
\frac{\gamma_{\rm tip}}{1+\alpha^2}
{\rm sgn}\!\left[E_{\rm S}(\phi)\right]
\sin\phi\,,
\end{equation}
again up to the normalization convention discussed above. Away from the
parity-changing point, one may replace
\({\rm sgn}[E_{\rm S}(\phi)]\) by \({\rm sgn}(1-\alpha^2)\), which represents the weak-tunneling expression. However, near the crossing, the dressed
energy \(E_{\rm S}(\phi)\) must be kept within the sign function, as the tip
can shift the Shiba pole through zero energy.

\subsection{Relation between the projected and microscopic tunneling scales}

The effective tunneling amplitude \(t_{\rm S}\) in
\cref{eq:HamiltonianLowEenergy} is obtained by projecting the
microscopic tip--substrate tunneling amplitude \(t_{\rm tip}\) onto the
local Shiba wave function at the position of the STM tip. As discussed in
the main text, this gives
\[
t_{\rm S}=t_{\rm tip}\sqrt{W_{\rm S}}\,,
\]
where \(W_{\rm S}\) is the local Nambu weight of the Shiba state at the
control adatom.

This relation can be made explicit from the pole structure of the local
retarded Green's function. Close to the isolated Shiba pole (hence, in the absence of the tip),
\begin{equation}
g_S^R(\omega)
\simeq
\frac{Z_{\rm S}}
{\omega-\epsilon_{\rm S}+i0^+}\,,
\end{equation}
where \(Z_{\rm S}\) is the local spectral-weight matrix of the
Shiba state. The projected local weight \(W_{\rm S}\) is determined by the
corresponding electron and hole components of this residue. For the
classical-impurity Green's function used above, the residue scales as
\[
W_{\rm S}\sim Z_{\rm S}\sim \pi\nu_0\Delta\,,
\]
up to a dimensionless function of \(\alpha\), which remains of order unity
when the Shiba state is well separated from the quasiparticle continuum.
Consequently,
\begin{equation}
|t_{\rm S}|^2
=
|t_{\rm tip}|^2 W_{\rm S}
\sim
|t_{\rm tip}|^2\pi\nu_0\Delta\,.
\end{equation}

The tunneling rate pertaining to projected Shiba level induced by the superconducting tip is therefore
\begin{equation}
\Gamma_{\rm S}
=
\pi\nu_0 |t_{\rm S}|^2
\sim
(\pi\nu_0 t_{\rm tip})^2\Delta
=
\gamma_{\rm tip}\Delta\,.
\end{equation}
Thus, the weak-coupling condition in the projected low-energy Hamiltonian, $\Gamma_{\rm S}/\Delta\ll1$, is equivalent to $\gamma_{\rm tip}\ll1$. The superconducting gap enters the projected tunneling rate through the local Shiba residue \(W_{\rm S}\), while the microscopic full Green-function calculation is naturally organized in terms of dimensionless normal-state coupling \(\gamma_{\rm tip}\).

The Hamiltonian in \cref{eq:HamiltonianLowEenergy} corresponds to a project-first construction: the substrate is first projected onto the isolated Shiba and Majorana modes, and then the STM tip is coupled to these modes through the projected amplitudes \(t_{\rm S}\) and \(t_{\rm M}\). This description is controlled when the in-gap modes are well separated from the continuum and \(\Gamma_{\rm S}/\Delta\ll1\). For stronger tip coupling, the tip modifies the Shiba pole itself, including its energy, residue, and electron-hole composition. In that regime, one should first extract the dressed pole \(E_{\rm S}(\phi)\) and the corresponding local spinor from the full Dyson equation before constructing the low-energy Shiba--Majorana model. Using a fully dressed \(E_{\rm S}(\phi)\) together with an explicitly projected tip--Shiba tunneling term would double count the coupling to the tip. However, such a regime lies outside the scope of this work.

\section{Analytical derivation of the supercurrent in the low-energy limit} \label{Appendix:CurrentDerivation}
Here, we present the derivation of \cref{eq:SupercurrentExpressionZeroTSystem} using the perturbative Green’s function approach in the low-energy limit. Starting with the low-energy effective Hamiltonian in \cref{eq:HamiltonianLowEenergy}, first we define a basis with the Shiba-Majorana states $(\Psi_{\rm S})$ and the tip $(\Psi_{\rm T})$ as
\begin{equation}\label{eq:AppendixBasis}
    \begin{pmatrix}
    \Psi_{\rm S} & \Psi_{\rm T}
\end{pmatrix}^{T}, \Psi_{\rm S}=\begin{pmatrix}
    c^{\dagger} &
    c&
    d^{\dagger}&
    d
\end{pmatrix}^{\dagger}, \Psi_{\rm T}=\begin{pmatrix}
    a_{\uparrow}^{\dagger} &
    a_{\uparrow}&
    a_{\downarrow}^{\dagger}&
    a_{\downarrow}
\end{pmatrix}^{\dagger}, 
\end{equation}
where $c^{\dagger}, d^{\dagger}$ are the creation operators of the Majorana and control Shiba modes, respectively, and $a^{\dagger}_{\sigma}$ are the electron creation operators of the superconducting tip with spin $\sigma$. We write the matrix Hamiltonian (defined for the Shiba-Majorana state as $H_{\rm S}$, superconducting tip as $H_{\rm T}$ and the tunnel coupling as $H_{\rm V}$) and the Green's functions 
\begin{equation}\label{eq:AppendixMatrixH}
\begin{split}
   H=& \begin{pmatrix}
         H_\mathrm{S} & H_\mathrm{V}\\
          H^{\dagger}_V &  H_\mathrm{T}
    \end{pmatrix}, G= \begin{pmatrix}
         G_{SS} & G_{ST}\\
          G_{TS} &  G_{TT}
    \end{pmatrix},  g= \begin{pmatrix}
         g_{SS} &0\\
          0 &  g_{TT}
    \end{pmatrix}, \\\\
    H_{\mathrm{S}}=&\dfrac{1}{2}\begin{pmatrix}
        \epsilon_\mathrm{M} & 0 & t_1^* & t^*_2 \\
        0 & -\epsilon_\mathrm{M} & -t_2 & -t_1 \\
        t_1 & -t_2^* & \epsilon_\mathrm{ S} & 0\\
        t_2 & -t_1^* & 0 & -\epsilon_\mathrm{ S}
    \end{pmatrix}, 
    \\
    H_{\mathrm{T}}=&\dfrac{1}{2}\begin{pmatrix}
        \epsilon_k & 0 & 0 & \Delta \\
        0 & -\epsilon_k & -\Delta & 0 \\
        0 & -\Delta &\epsilon_k & 0\\
       \Delta& 0 & 0 & -\epsilon_k
    \end{pmatrix}, 
    \\ 
     H_{\mathrm{V}}=&\dfrac{1}{2}\begin{pmatrix}
       0 & 0 & 0 & 0\\
0 & 0 & 0 & 0\\
t_{\rm S} e^{\frac{i\phi_1}{2}}  & 0 & 0 & -t_{\rm S} e^{\frac{-i\phi_1}{2}} \\
0  & -t_{\rm S} e^{\frac{-i\phi_1}{2}} &  t_{\rm S} e^{\frac{i\phi_1}{2}} & 0
    \end{pmatrix},
\end{split}
\end{equation}
where $g$ refers to the Green's function in the absence of $H_\mathrm{V}$ and $G$ is the full Green's function of the hybridized system. The bare Green's function related to $H_\mathrm{S}$ can be computed by means of Green's function equations of motion 
\begin{equation} \label{eq:AppendixgFss}
\begin{split}
    g_{\rm SS}=&\dfrac{1}{(\omega^2-E^2_\mathrm{ M})(\omega^2-E^2_\mathrm{ S})} \times\\&\begin{pmatrix}
        g_{SS,11} & g^*_{SS,21} & g^*_{SS,31} & g_{SS,14} \\
       g_{SS,21} & g_{SS,22} & g_{SS,23}  & g_{SS,24} \\
        g_{SS,31} & g^*_{SS,23} &g_{SS,33} & g^*_{SS,43} \\
        g^*_{SS,14}  & g^*_{SS,24} & g_{SS,43} & g_{SS,44}
    \end{pmatrix},   
\end{split}
\end{equation}
where the poles are 
\begin{equation}\label{eq:AppendixPoles}
\begin{split}
    E_\mathrm{ M}=& \dfrac{1}{2}\left(\sqrt{(\epsilon_\mathrm{M}+\epsilon_\mathrm{ S})^2+4|t_2|^2}-\sqrt{(\epsilon_\mathrm{M}-\epsilon_\mathrm{ S})^2+4|t_1 |^2}\right),\\
      E_\mathrm{ S}=& \dfrac{1}{2}\left(\sqrt{(\epsilon_\mathrm{M}+\epsilon_\mathrm{ S})^2+4|t_2|^2}+\sqrt{(\epsilon_\mathrm{M}-\epsilon_\mathrm{ S})^2+4|t_1|^2}\right),
    \end{split}
\end{equation}
the off-diagonal matrix elements are
\begin{equation} \label{eq:AppendixDefinitionsOffDiagonal}
\begin{split}
     &g_{\rm{SS},21}=-2\omega t_1 t_2, \ g_{\rm{SS},43}= 2 \omega t^*_1 t_2,\\
    &g_{\rm{SS},14} =t^*_2 \left[|t_1|^2-|t_2|^2 +(\omega-\epsilon_\mathrm{ S})(\omega+\epsilon_\mathrm{ M}) \right],\\ &g_{\rm{SS},24}=t_1 \left[|t_1|^2-|t_2|^2 -(\omega-\epsilon_\mathrm{ S})(\omega-\epsilon_\mathrm{ M}) \right], \\
    &g_{\rm{SS},23}=t_2 \left[|t_2|^2-|t_1|^2 -(\omega+\epsilon_\mathrm{ S})(\omega-\epsilon_\mathrm{ M}) \right], \\
    &g_{\rm{SS},31}=t^*_1 \left[|t_2|^2-|t_1|^2-(\omega+\epsilon_\mathrm{ M})(\omega+\epsilon_\mathrm{ S})\right],
 \end{split}
\end{equation}
and diagonal matrix elements are
\begin{equation} \label{eq:AppendixDefinitionsDiagonal}
\begin{split}
   &g_{\rm{SS},11}=(\omega^2-\epsilon_\mathrm{ S}^2)(\omega+\epsilon_\mathrm{M})-(\omega-\epsilon_\mathrm{ S})|t_1|^2 -(\omega+\epsilon_\mathrm{ S})|t_2|^2, \\
   &g_{\rm{SS},22}=(\omega^2-\epsilon_\mathrm{ S}^2)(\omega-\epsilon_\mathrm{M})-(\omega-\epsilon_\mathrm{ S})|t_2|^2 -(\omega+\epsilon_\mathrm{ S})|t_1|^2,\\
   &g_{\rm{SS},33}=(\omega^2-\epsilon_\mathrm{M}^2)(\omega+\epsilon_\mathrm{ S})-(\omega-\epsilon_\mathrm{M})|t_1|^2 -(\omega+\epsilon_\mathrm{M})|t_2|^2 ,\\
    &g_{\rm{SS},44}=(\omega^2-\epsilon_\mathrm{M}^2)(\omega-\epsilon_\mathrm{ S})-(\omega-\epsilon_\mathrm{M})|t_2|^2 -(\omega+\epsilon_\mathrm{M})|t_1|^2.\\
    \end{split}
\end{equation}
The superconducting tip's Green's function is defined as,
\begin{equation} \label{eq:AppendixTipGF}
g_{\rm{TT}}=\dfrac{-\pi \nu_0}{ \sqrt{\Delta^2-\omega^2}}
    \begin{pmatrix}
        \omega & 0 &0 & \Delta \\
        0 & \omega & -\Delta & 0 \\
        0 & -\Delta &\omega &0 \\
        \Delta & 0 & 0 &\omega
    \end{pmatrix}.
\end{equation}
Next, we rewrite the full Green's function in terms of the self-energy $\Sigma$,
\begin{equation}\label{eq:AppendixSelfEnergy}
\begin{split}
        G=\dfrac{g}{1-g\Sigma},\ 
     \Sigma= \begin{pmatrix}
         0 & H_\mathrm{V}\\
      H^{\dagger}_{\rm{V}} & 0
    \end{pmatrix}\,.
\end{split}
\end{equation}
The component $G_\mathrm{SS}$, defined in \cref{eq:AppendixMatrixH}, can be explicitly written as 
\begin{equation}\label{eq:AppendixGFSS}
       G_{\rm{SS}}=\dfrac{g_{\rm{SS}}}{1-g_{\rm{SS}}H_\mathrm{V}g_{\rm{TT}}H^{\dagger}_{\rm{V}}}\,.
\end{equation}
The poles in \cref{eq:AppendixGFSS} refer to the modified eigenstates in the presence of the tip, $\tilde{E}_\mathrm{M},\tilde{E}_\mathrm{S}$, and can be acquired by imposing the condition 
\begin{equation}\label{eq:AppendixDeterminantCondition}
    \det(1-g_{\rm{SS}}H_\mathrm{V}g_{\rm{TT}}H^{\dagger}_{\rm{V}})=0\,.
\end{equation}
Combining \cref{eq:AppendixgFss,eq:AppendixMatrixH,eq:AppendixTipGF},
we obtain the following equation for the subgap poles 
\begin{widetext}
\begin{equation}\label{eq:AppendixCondition}
    \begin{split}
          &A_0(\omega)+A_1(\omega)+A_2(\omega)=0,\\
          &A_0(\omega)=\left(\Delta^2-\omega^2\right)(\omega^2-E^2_\mathrm{\rm M})^2(\omega^2-E^2_\mathrm{\rm S})^2,\\
          &A_1(\omega)=\Gamma_{\rm S}\sqrt{\Delta^2-\omega^2}(\omega^2-E^2_\mathrm{\rm M})(\omega^2-E^2_\mathrm{ S})\left(\omega^2\left(\omega^2-\epsilon^2_{\rm{M}}-\abs{t_1}^2-\abs{t_2}^2\right)-\Delta\cos(\phi)\left(\epsilon_\mathrm{ S}\left(\omega^2-\epsilon^2_{\rm{M}}\right)+\epsilon_\mathrm{M}\left(\abs{t_1}^2-\abs{t_2}^2\right)\right)\right),\\
          &A_2(\omega)=\dfrac{2\Gamma_{\rm S}^2\left(\omega^2-\epsilon^2_{\rm{M}}\right)\left(\omega^2-\Delta^2\cos^2(\phi)\right)}{4}  ( \left(\omega^2-\epsilon^2_{\rm{M}}\right)\left(\omega^2-\epsilon^2_{\rm{S}}\right)-2\omega^2\left(\abs{t_1}^2+\abs{t_2}^2\right)
    \\
    & \hspace{8ex} -2\epsilon_\mathrm{M}\epsilon_\mathrm{ S}\left(\abs{t_1}^2-\abs{t_2}^2\right)+\left(\abs{t_1}^2-\abs{t_2}^2\right)^2 )\,,
    \end{split}
\end{equation}
\end{widetext}
    where we have split the contributions into powers of $\Gamma_{\rm S}$. The condition \cref{eq:AppendixCondition} is now treated perturbatively in the powers of $\Gamma_{\rm S}$. We consider the expansions $\tilde{E}_\mathrm{ M}=E_\mathrm{ M}+\Gamma_{\rm S} x$ and $\tilde{E}_\mathrm{ S}=E_\mathrm{ S}+\Gamma_{\rm S} y$, solve \cref{eq:AppendixCondition} for $x$ and $y$ in the zeroth order in $\Gamma_S$ to obtain the following  
\begin{widetext}
\begin{equation}\label{eq:AppendixXY}
    \begin{split}
           &x=-\dfrac{E_{\rm{M}}^2\left(E_{\rm{M}}^2-\epsilon^2_{\rm{M}}-\abs{t_1}^2-\abs{t_2}^2\right)-\Delta\cos(\phi)\left(\epsilon_\mathrm{ S}\left(E_{\rm{M}}^2-\epsilon^2_{\rm{M}}\right)+\epsilon_\mathrm{M}\left(\abs{t_1}^2-\abs{t_2}^2\right)\right)}{2\sqrt{\Delta^2-E_{\rm{M}}^2}E_{\rm{M}}(E_{\rm{M}}^2-E^2_{\rm{S}})}\,,\\
             &y=-\dfrac{E_{\rm{S}}^2\left(E_{\rm{S}}^2-\epsilon^2_{\rm M}-\abs{t_1}^2-\abs{t_2}^2\right)-\Delta\cos(\phi)\left(\epsilon_\mathrm{ S}\left(E_{\rm{S}}^2-\epsilon^2_{\rm M}\right)+\epsilon_\mathrm{M}\left(\abs{t_1}^2-\abs{t_2}^2\right)\right)}{2\sqrt{\Delta^2-E_{\rm{S}}^2}E_{\rm{S}}(E_{\rm{S}}^2-E^2_{\rm{M}})}\,,
    \end{split}
\end{equation}
\end{widetext}
where $|E_\mathrm{M}|\neq |E_\mathrm{S}|$ is assumed. Note that only the terms proportional to $\cos{\phi}$ are relevant for the supercurrent. For the low-energy states, the Bogoliubov form of the Hamiltonian is
\begin{equation}\label{eq:AppendixBogoHam}
    H_\mathrm{ B}= |\tilde{E}_{\rm{M}}|\left(f_1^{\dagger}f_1-\frac{1}{2}\right)+|\tilde{E}_{\rm{S}}|\left(f_2^{\dagger}f_2-\frac{1}{2}\right)\,,
\end{equation}
where the $f^{\dagger}_1,f^{\dagger}_2$ operators fill the states with energies $\tilde{E}_\mathrm{ M}$ and $\tilde{E}_\mathrm{ S}$. Here, the ground-state energy is identified by $E_\mathrm{GS}=-\left(|\tilde{E}_\mathrm{ M}|+|\tilde{E}_\mathrm{ S}|\right)/2$. 
The average supercurrent can therefore be computed as the derivative of the free energy $F=-k_B T {\rm ln} Z$, where  $Z$ is the partition function, 
\begin{equation}\label{eq:CurrGeneral}
    \mathcal{I}= \dfrac{\partial}{\partial \phi} (-k_B T {\rm ln} Z )\,,
\end{equation}
and $k_B$ is the Boltzmann constant. For the specific form of the two-level Bogoliubov Hamiltonian in \cref{eq:AppendixBogoHam}, the average supercurrent becomes
\begin{equation} \label{eq:AppendixMajSuperFormalSuperCurrent}
    \mathcal{I}=-\tanh(\dfrac{|\tilde{E}_\mathrm{M}|}{2k_BT})\partial_{\phi}x-\tanh(\dfrac{|\tilde{E}_\mathrm{S}|}{2k_BT})\partial_{\phi}y\,.
\end{equation}
In the case of a large superconducting gap, $\Delta\gg E_\mathrm{M,S}$ and zero-temperature, $T\rightarrow0$, the expression in \cref{eq:AppendixMajSuperFormalSuperCurrent} simplifies to the expression in \cref{eq:SupercurrentExpressionZeroTSystem}.
\section{Supercurrent calculation from Keldysh Green's function formalism: a numerical analysis}\label{Appendix:KeldyshMethod}
The numerical Keldysh mechanism acts as a tool to investigate the supercurrent from the superconducting tip to the Shiba-Majorana hybridized state beyond the weak-coupling theory. Starting with the low-energy Hamiltonian in \cref{eq:HamiltonianLowEenergy}, we calculate the supercurrent from the superconducting tip to the adatom using the current formulation 
$\mathcal I=i\left\langle\left[\mathcal N_{\rm tip}, H_{\rm tip-low}\right]\right\rangle$ 
as stated in \cref{Model}. It is important to mention that this generic current formula leads to the same current expression as obtained by \cref{eq:SupercurrentExpressionZeroT} or \cref{eq:CurrGeneral} by calculating the commutation relation, using the model Hamiltonian~\cite{Mucciolo2025}. We define the Keldysh Green's function in the Bogoliubov–de Gennes (BdG) representation~\cite{Sun1999, Zhu2002, Cheng2023, Sun2023, Debnath2024, Debnath2025}
\begin{equation}
{\mathcal{G}}^{<}(\omega) \equiv \left\langle \left( 
\begin{array}{l}
\Psi_{\rm{S}} \\ 
\Psi_{\rm{T}} 
\end{array}
\right)^{\dagger} \otimes \left( 
\begin{array}{ll}
\Psi_{\rm{S}} & \Psi_{\rm{T}}
\end{array} 
\right) \right\rangle\,,
\end{equation}
and considering the basis in \cref{eq:AppendixBasis}, the Keldysh lesser Green's function for the Hamiltonian in \cref{eq:HamiltonianLowEenergy} can be written as~\cite{Cheng2023, Sun2023, Debnath2024, Debnath2025}
\begin{eqnarray}
\mathcal{G}^{<}=\left(\begin{array}{ccc}
\mathcal{G}_{\text{SS}}^{<}&\mathcal{G}_{\text{ST}}^{<}\\
\mathcal{G}_{\text{TS}}^{<}&\mathcal{G}_{\text{TT}}^{<}
\end{array}\right),
\label{eq:Greenless}
\end{eqnarray}
where $\mathcal{G}^{<}$ is an $8\cross 8$ matrix, $\mathcal{G}_{\text{SS}}^{<}$ and $\mathcal{G}_{\text{TT}}^{<}$ are the Green's functions for the Shiba state and the tip, and $\mathcal{G}_{\text{ST(TS)}}^{<}$ represents the tunneling Green's function. With this definition, the resulting supercurrent between the tip and the SC at the adatom location is given by
\begin{equation}
\begin{split}
\mathcal{I} &= 2 \int \frac{d\omega}{2\pi} \Big[
t_{\rm S} e^{\frac{i\phi_1}{2}} \left(\mathcal{G}^{<}_{34}(\omega)+\mathcal{G}^{<}_{42}(\omega)\right) \\
&\quad + t_{\rm M} e^{\frac{i\phi_2}{2}} \left(\mathcal{G}^{<}_{14}(\omega)+\mathcal{G}^{<}_{22}(\omega)\right)
\Big]\,.
\end{split}
\label{eq:I}
\end{equation}
where, $\mathcal{G}^{<}_{ij}(\omega)$ represents the $\langle i,j\rangle$--th component of $\mathcal{G}^{<}(\omega)$. To calculate this Keldysh lesser Green's function, we use the fluctuation-dissipation theorem at thermal equilibrium~\cite{Kubo1966, Kamenev2005, Kamenev2009}
\begin{eqnarray} 
    \mathcal{G}^{<}(\omega)&=&-f(\omega)(\mathcal{G}^{\text{r}}-\mathcal{G}^{\text{a}})+(1-f(\omega))(\mathcal{G}^{\text{r}}-\mathcal{G}^{\text{a}})\nonumber\\
    ~~~&=&\text{tanh}\left(\frac{\omega}{2 k_{B}T}\right) (\mathcal{G}^{\text{r}}-\mathcal{G}^{\text{a}})\,,
    \label{eq:FDT}
\end{eqnarray}
where $f(\omega)$ is the Fermi function and $\mathcal{G}^{\text{r(a)}}(\omega)$ is the Keldysh retarded (advanced) Green's function, which we further numerically evaluate using the Dyson equation of motion~\cite{Dyson1949, Sun2000a, Villas2020, Cheng2023}
\begin{eqnarray}
    \mathcal{G}^{r}=g^{r}+g^{r} \Sigma^{r} \mathcal{G}^{r}\,.
    \label{eq:Dyson}
\end{eqnarray}
Here, $\Sigma^{\text{r}}$ is the self-energy of the system and $g^{\text{r}}$ is the non-interacting bare Green's function for the model Hamiltonian, which can be expressed as
\begin{eqnarray}
g^{\text{r}}=\left(\begin{array}{cc}
g_{\text{MM}}  & 0\\
0  &  g_{\text{TT}}
\end{array}\right)\,,
\label{eq:gr}
\end{eqnarray}
where $g_{\text{MM}}$ and $g_{\text{TT}}$ represent the Green's function for the uncoupled Shiba-Majorana state and the isolated superconducting tip, respectively. The tip's Green's function $g_{\text{TT}}$ is the same as in \cref{eq:AppendixTipGF}. To ensure numerical stability and avoid divergences, we introduce a small broadening parameter $\eta$ that shifts the energy as  $\omega\rightarrow\omega+i\eta$, which effectively accounts for finite lifetime effects of the tunneling quasiparticles. We have considered $\eta=10^{-5}$ throughout our numerical analysis, unless otherwise specified.

We derive the uncoupled Green's function for the Shiba-Majorana state as,
\begin{eqnarray}
g_{\text{MM}}=\left(\begin{array}{cccc}
\frac{1}{E-\epsilon_{\mathrm{M}}+i\eta}  & 0 & 0 & 0 \\
0  & \frac{1}{E+\epsilon_{\mathrm{M}}+i\eta} & 0 & 0\\
0 & 0 & \frac{1}{E-\epsilon_{\mathrm{S}}+i\eta}  & 0 \\
0 & 0 & 0  & \frac{1}{E+\epsilon_{\mathrm{S}}+i\eta} 
\end{array}\right).
\label{eq:g_MM}
\end{eqnarray}

The self-energy is
\begin{eqnarray}
\Sigma^{\text{r}}=\left(\begin{array}{cc}
V_{\text{MM}} & V_{\text{MT}}\\
V_{\text{TM}}  &  0
\end{array}\right),
\label{eq:AppendixSelfenergy}
\end{eqnarray}
with $V_{\text{MT}}=V^{\dagger}_{\text{TM}}$, which signifies the tunnel coupling between the tip--Shiba and tip--Majorana states
\begin{equation}
   V_{\mathrm{MT}}=\frac{1}{2}\begin{pmatrix}
       t_{\rm M} e^{\frac{i\phi_2}{2}} & 0 & 0 & - t_{\rm M} e^{\frac{-i\phi_2}{2}}\\
0 & - t_{\rm M} e^{\frac{-i\phi_2}{2}} &  t_{\rm M} e^{\frac{i\phi_2}{2}} & 0\\
 t_{\rm S} e^{\frac{i\phi_1}{2}}  & 0 & 0 & -t_{\rm S} e^{\frac{-i\phi_1}{2}} \\
0  & -t_{\rm S} e^{\frac{-i\phi_1}{2}} &  t_{\rm S} e^{\frac{i\phi_1}{2}} & 0
    \end{pmatrix}.
\end{equation}
The hybridization of the Shiba state to the Majoranas enters the self-energy term as~\cite{Sun2000a, Cheng2023, Sun2023, Debnath2024, Mucciolo2025, Debnath2025} 
\begin{eqnarray}
V_{\rm MM}=\left(\begin{array}{cccc}
0 & 0 & t^{*}_{1} & t^{*}_{2}\\
0 & 0 & -t_{2} & -t_{1}\\
t_{1}  & -t^{*}_{2} & 0 & 0 \\
t_{2}  & -t^{*}_{1} &  0 & 0
\end{array}\right).
\end{eqnarray}
Given the non-interacting retarded Green's function in \cref{eq:gr} and self-energy in \cref{eq:AppendixSelfenergy}, we use the Dyson equation in \cref{eq:Dyson} and the fluctuation-dissipation theorem in \cref{eq:FDT}, to calculate the total retarded Green's function and compute the supercurrent as $\mathcal{I}/\Gamma$ using \cref{eq:I}, where $\Gamma=\Gamma_{\rm S}+\Gamma_{\rm M})$ is the tunneling rate of supercurrent from the tip to the Shiba-Majorana hybridized state as defined in \cref{eq:Gamma_SM} of the main text. The electronic density of states shown in \cref{fig:2}(d) is defined as~\cite{Picano2021}
\begin{equation}\label{eq:dos}
    \rho(\omega)=-\frac{1}{\pi} \mathrm{Tr}[\mathrm{Im}[\mathcal{G}^{r}(\omega)]].
\end{equation}
\section{Additional results}
\subsection{Effect of particle-hole asymmetry and bare Shiba energy on the supercurrent driven by tip--Shiba coupling}\label{Appendix:Particle-hole asymmetry}
\begin{figure}
    \centering
    \includegraphics[width=1\linewidth,height=0.24\textwidth]{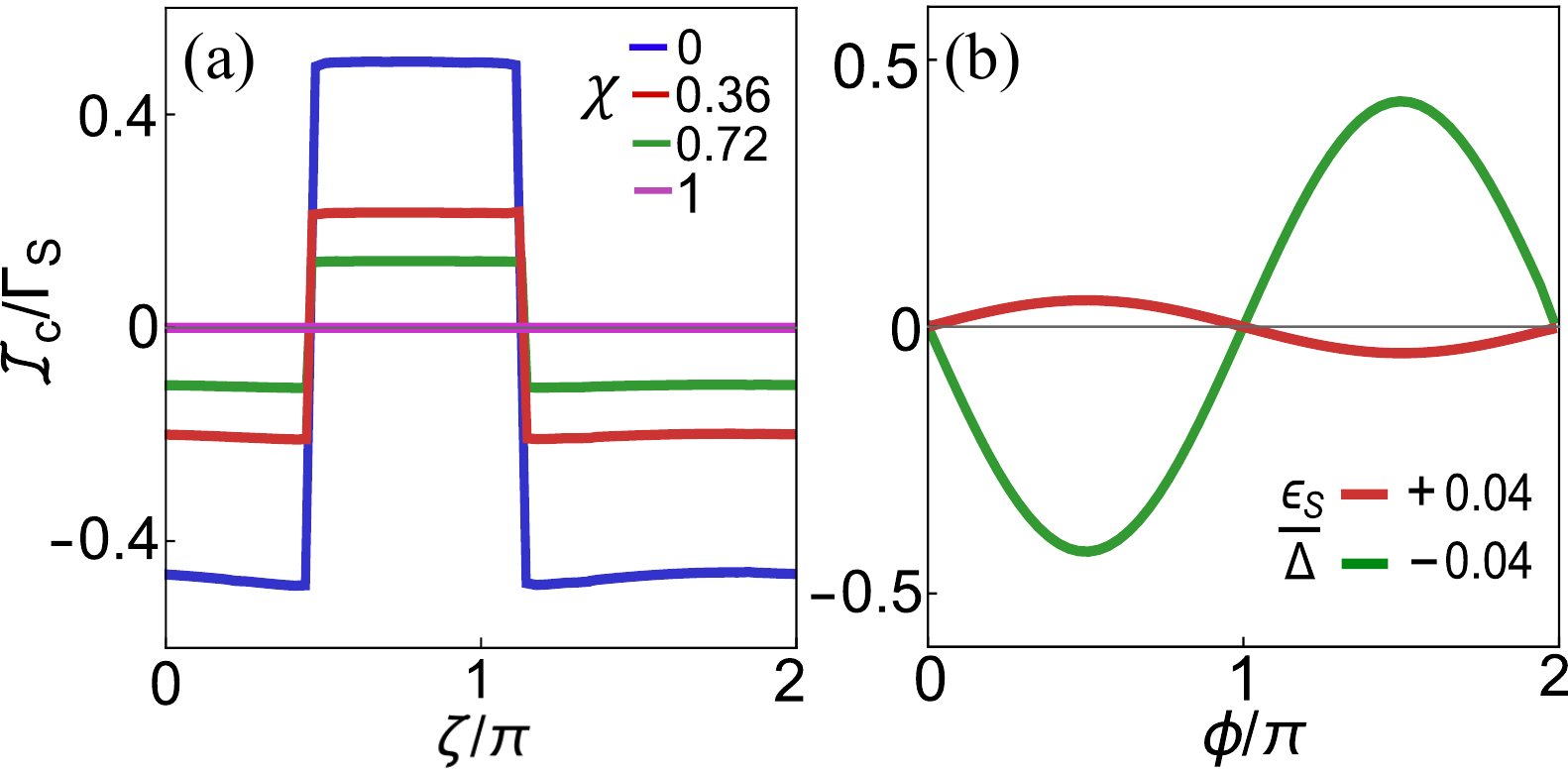}
    \caption{Effect of particle-hole asymmetry to the Shiba state and adatom-Shiba energy on supercurrent. (a) Supercurrent with respect to the adatom magnetization angle $\upzeta$ shows the effect of particle-hole asymmetry $\chi=u-h$ for different $u$ and $h$ values. All other parameters are considered the same as in \cref{fig:2}(b). The asymmetry in particle-hole state gradually reduces the supercurrent. (b) Tip--to--Shiba supercurrent for positive and negative Shiba energies at particle-hole symmetric regime, showing a $0$--$\pi$ transition when the sign of $\epsilon_{\mathrm S}$ is reversed. The parameters are $\Gamma_{\rm S}/\Delta=10^{-4}, \Gamma_{\rm M}/\Delta=0$, $\epsilon_{\mathrm M}=-0.02\Delta$, and $(\upzeta,\theta)=(\pi/2,\pi/2)$.}
\label{fig:6}
\end{figure}
\begin{figure*}
    \centering
    \includegraphics[width=\linewidth,height=0.27\textwidth]{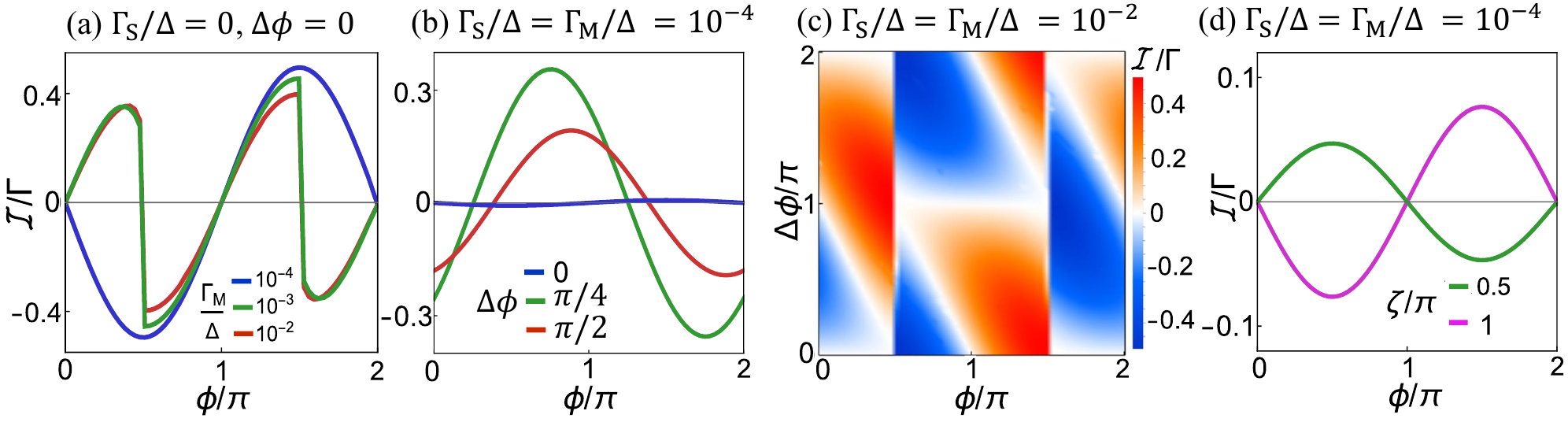}
    \caption{Supercurrent in presence of finite tip-to-Majorana coupling. (a) Supercurrent with respect to the superconducting phase $\phi$ for different tip--Majorana tunneling rates $\Gamma_\mathrm{M}/\Delta$ at tip--Shiba tunneling rate $\Gamma_\mathrm{S}/\Delta=0$. (b) Effect of additional superconducting phase $\Delta\phi$ associated to the tip--Majorana coupling in the total supercurrent at $\Gamma_{\rm S}/\Delta=\Gamma_{\rm M}/\Delta=10^{-4}$ with respect to the superconducting phase $\phi$, where $\Gamma=(\Gamma_{\rm S}+\Gamma_{\rm M})$. (c) 2D plot of supercurrent in the plane of $\phi-\Delta\phi$ at $\Gamma/\Delta=10^{-2}$. In (a)-(c) Majorana and Shiba bare energies are considered as $\epsilon_\mathrm{M}=-0.2\Delta$ and $\epsilon_\mathrm{S}=0.002\Delta$ and the adatom magnetization angles are $(\upzeta,\theta)=(\pi/2,\pi/2)$. (d) The sign reversal of supercurrent with respect to the superconducting phase $\phi$ for different adatom magnetization angle $\upzeta$ presents the signature of quantum phase transition in the Majorana-Shiba hybridized state in the presence of both tip--Shiba and tip--Majorana couplings. Here we fix $\Delta\phi=0$, $\theta=\pi/2$, $\epsilon_\mathrm{M}=-0.02\Delta, \epsilon_\mathrm{S}=0.04\Delta$ and $\Gamma_{\rm S}/\Delta=\Gamma_{\rm M}/\Delta=10^{-4}$.}
\label{fig:7}
\end{figure*}
The tunneling Hamiltonian $H_\mathrm{tip-low}$ in \cref{eq:HamiltonianLowEenergy} contains the particle-hole weights through the terms $u$ and $h$. Throughout the main text, we  consider equal particle and hole weights, i.e., $u=h$. In this section, we investigate the influence of particle-hole asymmetry, i.e., considering $u\neq h$. Considering different pairs of $u$ and $h$ values, we define the particle-hole asymmetry parameter $\chi=\abs{u-h}$ and calculate the supercurrent for finite tip--Shiba coupling $\Gamma_\mathrm{S}/\Delta$ at tip--Majorana coupling $\Gamma_\mathrm{M}/\Delta=0$. We show the supercurrent in \cref{fig:6}(a) with respect to the adatom magnetization angle $\upzeta$ for different combinations of $u$ and $h$ values. The jump in the supercurrent at critical magnetization angles $\upzeta_{c1}$ and $\upzeta_{c2}$ remains unchanged at $\chi \neq 0$ compared to the symmetric particle-hole state with $\chi=0$, which signifies the jump in supercurrent is a robust characteristics of the quantum phase transition. As the particle-hole asymmetry diverts the spectral weight of the tunneling away from the coherent Copper pair transfer, it reduces the effective tunneling rate of the Josephson junction. As a result, the supercurrent strongly depends on the particle-hole asymmetry. 

In \cref{fig:6}(b), we plot the supercurrent in dependence on the superconducting phase $\phi$ at $\Gamma_{\rm M}/\Delta=0$. Though in the presence of finite tip--Majorana coupling, an additional phase $\Delta\phi$ modulates the current profile with the sign of bare energy of the Shiba state, as observed in \cref{fig:4}(c), we observe a clear signature of a `$0-\pi$' phase transition with a sign change of $\epsilon_{\rm S}$ in the presence of only tip--Shiba coupling. 
\subsection{Supercurrent in the presence of finite tip--Majorana coupling} \label{Appendix:two phases}
As described in \cref{TipMajoranaCoupling}, in a realistic setup, a tip--Majorana coupling with an additional complex phase $\Delta\phi$ as described in \cref{Model} could become relevant. To study the effect of this additional tunnel coupling $t_{\rm M}$ on the supercurrent, we first calculate the supercurrent in the tip--Majorana coupled regime by numerically fixing $\Gamma_\mathrm{S}/\Delta=0$. In \cref{fig:7}(a), we show that when the tip--Majorana tunneling rate $\Gamma_\mathrm{M}/\Delta$ is finite, the supercurrent with respect to the superconducting phase $\phi$ shows the regular Josephson-like sinusoidal in the weak tunneling limit. As we increase $\Gamma_\mathrm{M}/\Delta$, a sign change in supercurrent associated to zero-energy crossing is observed, similar to \cref{fig:2}(c).

To evaluate how the additional phase modifies the supercurrent, we plot the supercurrent in \cref{fig:7}(b), with respect to $\phi$ at different $\Delta\phi$. The figure also presents the line-cuts of \cref{fig:4}(b) at different $\Delta\phi$. We observe a shift in current--phase characteristics introduced by the additional tunneling phase associated to $\Gamma_{\rm M}/\Delta$. Next, we plot the supercurrent with respect to the superconducting phase $\phi$ and the additional phase $\Delta\phi$ for the tunneling rate $\Gamma/\Delta=10^{-2}$. We observe that the supercurrent profile changes qualitatively and quantitatively from the smooth interference pattern of the weak-tunneling limit, shown in \cref{fig:4}(b), and develops sharp phase boundaries at a strong tunneling rate. This indicates that stronger tip couplings hybridize the Shiba-Majorana spectrum more, such that phase-driven level rearrangements produce abrupt reversals of the Josephson current, as we observe for the supercurrent in strong tunneling regimes, driven by the zero-energy crossing of the total energy of the system. Hence, the resulting response is no longer as simple as we analytically obtained in \cref{eq:curr_delphi}, but is governed by the nontrivial phase evolution of the hybrid bound states and needs to be resolved numerically.

The adatom magnetization angles $(\upzeta, \theta)$ are intrinsically coupled to the additional superconducting phase $\Delta\phi$ through the superconducting tip. Therefore, the strongly hybridized Shiba–Majorana state precludes a simple analytical separation of the phase and the angular degrees of freedom, making a direct analysis of the current as a function of $\upzeta$ infeasible. To circumvent this, we fix $\Delta\phi=0$, and evaluate the supercurrent as a function of $\phi$ for different values of $\upzeta$ at $\theta=\pi/2$. The resulting sign reversal of the supercurrent in $\upzeta$, see \cref{fig:7}(d), confirms that the detection of the quantum phase transition remains robust. A finite $\Delta\phi$ would only shift the critical transition angle, not change the general characteristics.
\bibliography{refs1}

\end{document}